\documentclass[preprint,showpacs,preprintnumbers,amsmath,amssymb,
floatfix]{revtex4}

\usepackage{graphicx}% Include figure files
\usepackage{dcolumn}% Align table columns on decimal point
\usepackage{bm}% bold math
\def\Id{{\rm 1\kern-.3em I}}

\begin{document}

\title{Color transparency and short-range correlations in exclusive
pion photo- and electroproduction from nuclei}

\author{W. Cosyn}
\email{Wim.Cosyn@UGent.be}
\author{M.C. Mart\'{\i}nez}
\altaffiliation{Present address: Dpto FAMN, Universidad Complutense de 
Madrid, E-28040 Madrid, Spain.}
\author{J. Ryckebusch}

\affiliation{Department of Subatomic and Radiation Physics,\\
 Ghent University, Proeftuinstraat 86, B-9000 Gent, Belgium}
\date{\today}

\begin{abstract}
A relativistic and quantum mechanical framework to compute nuclear
transparencies for pion photo- and electroproduction reactions is
presented.  Final-state interactions for the ejected pions and
nucleons are implemented in a relativistic eikonal approach.  At
sufficiently large ejectile energies, a relativistic Glauber model can
be adopted.  At lower energies, the framework possesses the flexibility
to use relativistic optical potentials.  The proposed model can
account for the color-transparency (CT) phenomenon and short-range
correlations (SRC) in the nucleus.  Results are presented for
kinematics corresponding to completed and planned experiments at
Jefferson Lab.  The influence of CT and SRC on the nuclear
transparency is studied.  Both the SRC and CT mechanisms increase the nuclear
transparency.  The two mechanisms can be clearly separated, though, as they
exhibit a completely different dependence on the hard scale parameter.  The
nucleon and
pion transparencies as computed in the relativistic Glauber approach are
compared with optical-potential and semi-classical calculations.  The
similarities in the trends and magnitudes of the nuclear transparencies indicate
that they are not subject to strong model dependencies.
\end{abstract}


\maketitle 

\section{Introduction} \label{sec:introduction}
A commonly used variable to map the transition from hadronic to
partonic degrees of freedom is the nuclear transparency.  For a given
reaction process, it is defined as the ratio of the cross section per
target nucleon to the one from a free nucleon.  Accordingly, the
nuclear transparency provides a measure of the attenuation effects of
the nuclear medium on the hadrons produced in some reaction.  A
phenomenon finding its roots in QCD is color transparency (CT).  It
predicts the reduction of final-state interactions (FSI) of the
produced hadron with the surrounding nuclear medium at sufficiently
high momentum transfer. Thereby, the hadron is created in a
point-like configuration (PLC) and propagates as a color singlet
through the nucleus before evolving to the normal hadron state.  If CT
effects were to appear at a certain energy, the nuclear transparency
would be observed to overshoot the predictions from traditional
nuclear physics expectations.

Measurements of nuclear transparencies in search of CT have been
carried out with the $A(p,2p)$
\cite{Carroll:1988rp,Mardor:1998,Leksanov:2001ui,Aclander:2004zm} and
$A(e,e'p)$ \cite{Garino:1992ca,Makins:1994mm,O'Neill:1994mg,
Abbott:1997bc,Garrow:2001di,Dutta:2003yt} reactions, $\rho$-meson
production \cite{Adams:1994bw,Airapetian:2002eh} and diffractive
dissociation of pions into di-jets \cite{Aitala:2000hc}. Nuclear
transparencies for the pion photoproduction process $\gamma n
\rightarrow \pi^- p$ in $^4\text{He}$ have been measured in Hall A at
Jefferson Laboratory (JLab)\cite{Dutta:2003mk}. A Hall C experiment
has extracted the nuclear transparency for the pion electroproduction
process $e p \rightarrow e' \pi^+ n$ in $^2\text{H}$, $^{12}\text{C}$,
$^{27}\text{Al}$, $^{63}\text{Cu}$ and $^{197}\text{Au}$
\cite{Clasie:2007gq}. Ref.~\cite{Larson:2006ge} reports calculations
in a semi-classical model for the latter electroproduction experiment.
In Ref. \cite{Cosyn:2006vm}, we introduced a relativistic and quantum
mechanical model for computing the nuclear transparencies for the pion
photoproduction reaction and compared its predictions to the
$^4\text{He}(\gamma,p\pi^-)$ data and results from a semi-classical
model developed by Gao, Holt and Pandharipande \cite{Gao:1996mg}. In
this paper, we outline the model in more detail and extend it to
electroproduction reactions.  The intranuclear attenuation which
affects the ejectiles (nucleons and pions) is modeled in terms of a
relativistic eikonal approach.  The bound-state wave functions are
obtained from a relativistic mean-field model. At sufficiently small
values for the de Broglie wavelength, we use a relativistic version of
Glauber multiple scattering theory. At wavelengths approaching the
range of the nucleon-nucleon and pion-nucleon interaction length, the
model offers the flexibility to use optical potentials for modeling
FSI mechanisms. Short-range correlations (SRC) induce local
fluctuations in the nuclear density.  These corrections beyond the
mean-field approach influence the intranuclear attenuation. The
corresponding changes in the nuclear transparencies have been studied
in great depth within the context of $A(e,e'p)$ reactions
\cite{Nikolaev:1993sj,Frankel:1994wn,Alvioli:2003hv}. In $A(\gamma,N \pi)$ and
$A(e,e'N \pi)$
processes, both the emerging nucleons and pions are subject to these density
fluctuations. The SRC are incorporated into our model through the
introduction of a well-chosen central correlation function which induces
density correlations into the final system. In our
procedure, the proper normalization of the wave functions is
guaranteed.

Section \ref{sec:formalism} of the paper presents the formalism used
to calculate the nuclear transparencies.  A factorized expression for the 
cross section is derived for $A(\gamma,N\pi)$ (\ref{subsec:photo}) and $A(e,e'N\pi)$
(\ref{subsec:electro}). Next, in Sect.~\ref{subsec:fsi} the framework
for computing the effects stemming from FSI are discussed. Thereby,
special attention is paid to a parametrization of the $\pi N$
scattering parameters which are required in Glauber calculations.  In
Sect.~\ref{subsec:ct} the incorporation of the CT phenomenon and SRC
is discussed.  The results of our numerical calculations are presented
in Sect.~\ref{sec:results}.  FSI effects are investigated and
transparency results are shown for the pion photo- and
electroproduction reactions from various target nuclei.  Our
conclusions are stated in Sect.~\ref{sec:conclusion}.

\section{Formalism}\label{sec:formalism}
In this section, the formalism used to describe $A(\gamma,N\pi)$ and
$A(e,e'N\pi)$ 
reactions is presented.

\subsection{Pion photoproduction}
\label{subsec:photo}

We use the following notations for the four-momenta in the lab frame:
$q^\mu (q,\vec{q})$ for the photon, $P^\mu_A (E_A,\vec{p}_A= \vec{0})$
for the target nucleus, $P^\mu_{A-1} (E_{A-1},\vec{p}_{A-1})$ for the
residual nucleus, $P^\mu_N (E_N,\vec{p}_N)$ and $P^\mu_\pi
(E_\pi,\vec{p}_\pi)$ for the ejected nucleon and pion.  The missing
momentum $\vec{p}_m$ is defined as $\vec{p}_m\equiv -\vec{p}_{A-1}
=\vec{p}_N + \vec{p}_\pi - \vec{q}$ and the outgoing nucleon has spin
$m_s$.  The fivefold differential cross section in the lab frame reads
\begin{equation}
\label{cross}
\frac{d^5 \sigma}{dE_\pi d\Omega_\pi d\Omega_N} = \frac{M_{A-1} m_N
p_\pi p_N}{4(2\pi)^5q E_A} f^{-1}_{rec} \overline{\sum_{fi}}
\left| 
\mathcal{M}_{fi}^{(\gamma,N\pi)}
\right|^2 \,,
\end{equation}
with the recoil factor given by
\begin{equation}\label{eq:recoil}
f_{rec}=\frac{E_{A-1}}{E_A}\left|
1+\frac{E_N}{E_{A-1}}\left(1+\frac{(\vec{p}_\pi - \vec{q})\cdot
\vec{p}_N }{p^2_N}\right)\right|\,, 
\end{equation}
and 
$\mathcal{M}_{fi}^{(\gamma,N\pi)}$ the invariant matrix element:
\begin{equation} \label{matrixel}
\mathcal{M}_{fi}^{(\gamma,N\pi)} = \langle P_\pi^{\mu}, P_N^{\mu} m_s,
P_{A-1}^{\mu} J_R M_R | \hat{\mathcal{O}} | q^{\mu}, P_A^{\mu} 0^+
\rangle \, ,
\end{equation}
where $J_R M_R$ are the quantum numbers of the residual nucleus.  We restrict
ourselves to processes with an even-even target nucleus $A$.

The wave functions for the bound nucleons are constructed in an
Independent Particle Model (IPM).  We use relativistic wave functions
from the Hartree approximation to the Walecka-model with the W1
parametrization \cite{Furnstahl:1996wv}.  For the sake of conciseness,
only the spatial coordinates of the nucleons are written throughout
this work.  The single-particle wave functions $\phi_\alpha$ adopt the
following form for a spherically symmetric nuclear potential
\cite{Greiner:Wave}:

\begin{equation}
\phi_\alpha(\vec{r}) 
\equiv \phi_{n\kappa m}(\vec{r},\vec{\sigma})=
\left[ \begin{array}{c}
i \frac{G_{n\kappa}(r)}{r}\mathcal{Y}_{\kappa m}(\Omega,\vec{\sigma})\\
-\frac{F_{n\kappa} (r)}{r}\mathcal{Y}_{-\kappa m}(\Omega,\vec{\sigma})
\end{array} \right]\,.
\end{equation}
Here, $n$ is the principal quantum number, $\kappa$ and $m$ denote the
generalized angular momentum quantum numbers.  The spin spherical harmonics
$\mathcal{Y}_{\pm\kappa m}$ are defined as:
\begin{align}
\mathcal{Y}_{\kappa m}(\Omega,\vec{\sigma}) &= \sum_{m_l m_s} \langle l m_l
\frac{1}{2} m_s | jm \rangle Y_{l m_l} (\Omega) \chi
_{\frac{1}{2}m_s}(\vec{\sigma})\,,\nonumber\\
\mathcal{Y}_{-\kappa m}(\Omega,\vec{\sigma}) &= \sum_{m_l m_s} \langle \bar{l}
m_l
\frac{1}{2} m_s | jm \rangle Y_{\bar{l} m_l} (\Omega) \chi
_{\frac{1}{2}m_s}(\vec{\sigma})\,,
\end{align}

\begin{equation*}
\text{with} \quad j = |\kappa| - \frac{1}{2}\,, \quad l = 
\left\{ \begin{array}{lr}
 \kappa, & \kappa>0\\
 -\kappa-1, & \kappa<0
 \end{array}\right. \,, \quad  \bar{l} =
\left\{ \begin{array}{lr}
 \kappa-1, & \kappa>0\\
 -\kappa, & \kappa<0 \end{array}\right. \,.
\end{equation*}
The ground-state wave function of the target nucleus $|P_A^\mu
0^+\rangle \equiv \Psi_A^{\text{g.s.}}(\vec{r}_1,\ldots,\vec{r}_A)$ is
obtained by fully anti-symmetrizing the product of the individual
nucleon wave functions $\phi_{\alpha}$.  We model the pion
photoproduction process by means of a contact interaction: the initial
nucleon, impinging photon, the ejected pion and nucleon, couple in a
single space-time vertex.  As the process can take place on any of the
nucleons in the target nucleus, we get the following general
expression for the corresponding photoproduction operator:
\begin{equation}\label{eq:operator}
\hat{\mathcal{O}}=\sum_{i=1}^A O_\mu(\vec{r}_i)\, .
\end{equation}
We assume that $\hat{\mathcal{O}}$ is exempted from medium
effects. This is a common assumption in nuclear and hadronic physics
and is usually referred to as the impulse or quasi-free approximation (IA).  
In the context of $A(e,e'p)$ reaction, for example, the impulse approximation provides a fair 
description of the data \cite{Ryckebusch:2003fc}.  It is also applied in the 
experimental analysis of Ref. \cite{Clasie:2007gq} and the model of Ref. \cite{Lee:1999kd}.  The
impinging photon with polarization $\lambda$ is represented by
\begin{equation}
A^\mu(\lambda,\vec{r}_i)=\epsilon^\mu(\lambda) e^{i\vec{q}\cdot\vec{r}_i}\,.
\end{equation}
Here, $\epsilon^\mu(\lambda)$ is the polarization four-vector of the photon.
The wave function of the ejected nucleon is written as 
\begin{equation}
|P_N^\mu m_s\rangle
\equiv
\Psi^{(+)}_{\vec{p}_N,m_s}(\vec{r}_i)=\hat{\mathcal{S}}_{N'N}^\dagger(\vec{r}
_i;\vec { r }
_1,\ldots,\vec{r}_{j\neq
i},\ldots,\vec{r}_A)u(\vec{
p}_N,m_s)e^{i\vec{p}_N\cdot\vec{r}_i} \, ,
\end{equation}
which is the product of a positive-energy Dirac plane wave
$\phi_{\vec{p}_N}$ and an operator $\hat{\mathcal{S}}_{N'N}^\dagger$.
This operator describes the attenuation of the ejected nucleon through
soft final-state interactions with the other nucleons.
The wave function for the ejected pion adopts a similar form as the nucleon
one, i.e. a plane wave convoluted with a FSI factor $\hat{\mathcal{S}}_{\pi
N}^\dagger$: 
\begin{equation} 
|P_\pi^\mu \rangle
\equiv \Phi^{(+)}_{\vec{p}_\pi}(\vec{r}_i)=\hat{\mathcal{S}}_{\pi
N}^\dagger(\vec{r}_i;\vec{r}
_1,\ldots,\vec{r}_{j\neq
i},\ldots,\vec{r}_A)e^{i\vec{p}_\pi\cdot\vec{r}_i}\,.
\end{equation}
The final A-nucleon wave function is constructed by
anti-symmetrizing $\Psi^{(+)}_{\vec{p}_N,m_s}$ with the wave function for the
residual nucleus $\Psi_{A-1}^{J_R,m_R}$:
\begin{multline}
|P_N^\mu m_s, P_{A-1}^\mu J_R M_R \rangle
\equiv \Psi_A^{\vec{p}_N,m_s}(\vec{r}_1,\ldots,\vec{r}_A) =\\
\hat{\mathcal{A}}\left[\hat{\mathcal{S}}_{N'N}^\dagger
(\vec{r}_1;\vec{r}_2,\ldots,\vec{r}_A)
u(\vec{p}_N,m_s)e^{i\vec{p}_N\cdot\vec{r}_1}
\Psi_{A-1}^{J_R,m_R}(\vec{r}_2,
\ldots,\vec{r}_A)\right]\,.
\end{multline}

\begin{figure}[htb]
\centering
\includegraphics[width=15cm]{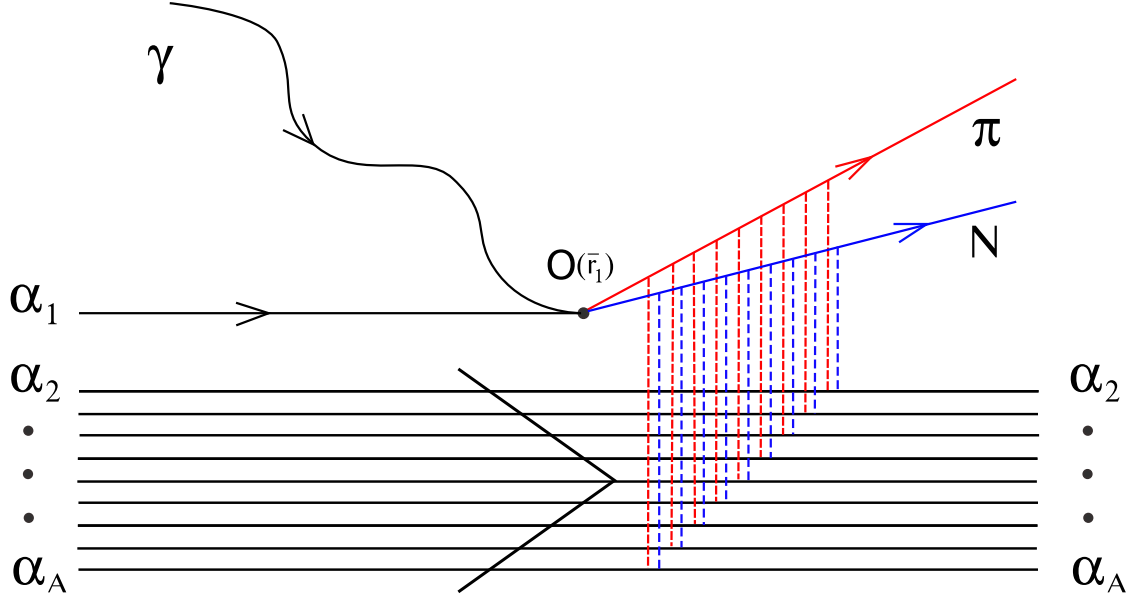}
\caption{(Color online) Diagram included in computing the matrix
element of Eq.~(\ref{timesA}).  The dashed lines denote the FSI of the
ejected pion (red) and nucleon (blue) with the spectator residual
nucleons. The diagram shown here is representative for the spectator
approximation: one active nucleon $N$ and $\pi$ are subject to soft
collisions with frozen spectator nucleons which occupy the single-particle
levels $\alpha_2,\alpha_3,\ldots,\alpha_A$ and are not subject to changes in
their quantum numbers.}
\label{fig:graph}
\end{figure}

As $\Psi_A^{\text{g.s.}}$ and $\Psi_A^{\vec{p}_N,m_s}$ are fully
anti-symmetric, each term of the operator (\ref{eq:operator}) will
yield the same contribution to the matrix element (\ref{matrixel}) and
we can restrict ourselves to the term with coordinate $\vec{r}_1$ and
multiply it with A.  With the above expressions for the operator and
the wave functions of the hadrons involved in the reaction, we can
write for the matrix element of Eq.  (\ref{matrixel}) in coordinate
space:
\begin{multline} \label{timesA}
\mathcal{M}_{fi}^{(\gamma,N\pi)}= A \int d\vec{r}_1 \int d\vec{r}_2 \ldots \int
d\vec{r}_A
\left[\Psi^{\vec{p}_N,m_s}_{A}\left(\vec{r}_1,\vec{r}_2,\ldots,\vec{r}
_{A}\right)\right]^\dagger\\
\times e^{-i\vec{p}_\pi\cdot\vec{r}_1}
\hat{\mathcal{S}}_{\pi N}(\vec{r}_1;\vec{r}_2,\ldots,\vec{r}
_A)O_\mu(\vec{r}_1) \epsilon^\mu(\lambda) 
e^{i\vec{q}\cdot\vec{r}_1}\Psi^{g.s.}_A\left(\vec{r}_1,\vec{r}_2,\ldots,\vec{r}_
{ A }\right)\,.
\end{multline}
We assume that $\hat{\mathcal{S}}_{N'N}$ and $\hat{\mathcal{S}}_{\pi N}$ are
spin independent and that only elastic and mildly inelastic collisions
with the spectator nucleons occur.  The actual nuclear transparency
measurements select events whereby the undetected final state with $(A-1)$
nucleons $ \left| P_{A-1}^{\mu} J_R
M_R \right>$ is left with little excitation energy, which makes these
assumptions very plausible. In computing the matrix element of
Eq.~(\ref{timesA}) we consider processes of the type displayed in
Fig.~\ref{fig:graph}. 
The following spectator approximation is assumed to be valid for a
struck nucleon with quantum numbers $\alpha_1$ :
\begin{align}\label{eq:approx}
\int d\vec{r}_1\ldots\int d\vec{r}_A \left[ 
\phi_{\vec{p}_N}(P_n(\vec{r}_1))\hat{\mathcal{S}}^\dagger_{N'
N}(P_n(\vec{r}_1);P_n(\vec{r}_2),\ldots,P_n(\vec{r}_A))
\phi_{\alpha_2}(P_n(\vec{r}_2))
\ldots\phi_{\alpha_A}(P_n(\vec{r}_A))
\right]^\dagger
\nonumber\\
\times e^{-i\vec{p}_\pi\cdot\vec{r}_1}\hat{\mathcal{S}}_{\pi
N}(\vec{r}_1;\vec{r}_2,\ldots,\vec{r}_A)
O_\mu(\vec{r}_1)e^{i\vec{q}\cdot\vec{r}_1}\phi_{\alpha_1}(P_m(\vec{r}_1))\phi_{
\alpha_2}(P_m(\vec{r}_2))\ldots\phi_{\alpha_A}(P_m(\vec{r}
_A))\nonumber\\
\approx \delta_{P_n(\vec{r}_2){P_m(\vec{r}_2)}}\ldots
\delta_{P_n(\vec{r}_A){P_m(\vec{r}_A)}} \int d\vec{r}_1 \ldots\int
d\vec{r}_A
\phi^\dagger_{\vec{p}_N}(\vec{r}_1)\hat{\mathcal{S}}_{N'N}
\left(\vec{r}_1;P_n(\vec{r}_2),\ldots,P_n(\vec{r}_A)\right)
\nonumber\\
e^{-i\vec{p}_\pi\cdot\vec{r}_1}\hat{\mathcal{S}}_{\pi
N}(\vec{r}_1;\vec{r}_2,\ldots,\vec{r}_A)
O_\mu(\vec{r}_1)e^{i\vec{q}\cdot\vec{r}_1}
\phi_{\alpha_1}(P_m(\vec{r}_1))
|\phi_{\alpha_2}(P_m(\vec{r}_2))|^2\ldots|\phi_{\alpha_A}(P_m(\vec{r}
_A))|^2\,,
\end{align}
with $P_m$ and
$P_n$ permutations of the set $\{\vec{r}_1,\ldots,\vec{r}_A\}$ occurring in the
anti-symmetrization of the nucleon wave functions.  
Due to the presence of the delta functions, the rhs of Eq. (\ref{eq:approx}) is
non-vanishing under the condition that $P_m(\vec{r}_1)=\vec{r}_1$ and
$P_m(\vec{r}_i)=P_n(\vec{r}_i)$ for $i=2,..,A$. This
means that both the bound wave function $\alpha_1$ and the
ejected nucleon have the same spatial coordinate as the
operator, $\vec{r}_1$.    Moreover, all $(A-1)!$ permutations of the subset
$\{\vec{r}_2,\ldots,\vec{r}_A\}$ yield an identical rhs.  

Thus, after expanding the wave functions in Eq.~(\ref{timesA}) and
employing Eq.~(\ref{eq:approx}), we arrive at
\begin{align}
\mathcal{M}_{fi}^{(\gamma,N\pi)}\approx \frac{A(A-1)!}{A!}\int d\vec{r}_1 \int
d\vec{r}_2 \ldots \int d\vec{r}_A
\Biggl[ 
|\phi_{\alpha_2}(\vec{r}_2)|^2\ldots|\phi_{\alpha_A}(\vec{r}
_A)|^2 \nonumber\\
\times 
u^\dagger (\vec{p}_N,
m_s)\hat{\mathcal{S}}_{\pi
N}(\vec{r}_1;\vec{r}_2,\ldots,\vec{r}_A)\hat{\mathcal{S}}_{
N'N}(\vec{r}_1;\vec{r}_2,\ldots,\vec{r}_A)\epsilon^\mu(\lambda)
O_\mu(\vec{r}_1)e^{-i\vec{p}_m\cdot\vec{r}_1}\phi_{\alpha_1}(\vec{r}_1)
\Biggr] \,
.
\end{align}
We now define the FSI factor $\mathcal{F}_{\text{FSI}}(\vec{r})$:
\begin{equation}\label{glauberdef}
\mathcal{F}_{\text{FSI}}(\vec{r})= \int d\vec{r}_2 \ldots \int d\vec{r}_A
|\phi_{\alpha_2}(\vec{r}_2)|^2\ldots|\phi_{\alpha_A}(\vec{r}_A)|^2
\hat{\mathcal{S}}_{\pi N}(\vec{r};\vec{r}_2,\ldots,\vec{r}_A)\hat{\mathcal{S}}_{
N'N}(\vec{r};\vec{r}_2,\ldots,\vec{r}_A)\,,
\end{equation}
and write
\begin{equation} \label{matrixel2}
\mathcal{M}_{fi}^{(\gamma,N\pi)}\approx \int d\vec{r}_1
\mathcal{F}_{\text{FSI}}(\vec{r}_1)
u^\dagger(\vec{p}_N,
m_s)\epsilon^\mu(\lambda)
O_\mu(\vec{r}_1)e^{-i\vec{p}_m\cdot\vec{r}_1}\phi_{\alpha_1}(\vec{r}_1)\,.
\end{equation}
In what follows, we assume that the pion production operator acts on a
bound-state wave function
as a scalar (factorization assumption):
$O_\mu(\vec{r})\phi_{\alpha_1}(\vec{r})\equiv \mathcal{C} \phi_
{\alpha_1}(\vec{r})$.  With
\begin{equation}\label{phid}
\phi_{\alpha_1}^D(\vec{p})=\frac{1}{(2\pi)^{3/2}}\int d\vec{r}
e^{-i\vec{p}\cdot\vec{r}}\phi_{\alpha_1}(\vec{r})\mathcal{F}^{\text{
FSI}}(\vec { r } )\, ,
\end{equation}
we can write
\begin{equation} \label{matrixfinal}
\mathcal{M}_{fi}^{(\gamma,N\pi)}\approx (2\pi)^{3/2} u^\dagger(\vec{p}_N,m_s)
\epsilon^\mu(\lambda) O_\mu(\vec{r}_1) \phi_{\alpha_1}^D(\vec{p}_m)\,.
\end{equation}
When studying nuclear transparencies, it is convenient to factorize
the invariant matrix element such that it becomes a convolution of a
factor describing the elementary
pion photoproduction process and a factor modeling the combined effect
of all FSI mechanisms
of the outgoing hadrons.  To reach this goal we relate the $\gamma +
A \rightarrow (A-1) +N+\pi$ matrix element in Eq. (\ref{matrixfinal})
to the one for for free nucleons $\gamma + N_i \rightarrow N + \pi$
\begin{equation}
\left(\mathcal{M}_{fi\,\mathrm{free}}^{(\gamma,N\pi)}\right)_{m_s,m'_s}=
u^\dagger(\vec{p}_N,m_s) \epsilon^\mu(\lambda) O_\mu(\vec{r}_1)
u(\vec{p}_m, m_{s'})\,,
\end{equation}
with $m_{s'}$ the spin of the initial nucleon.  First, we consider the
situation with vanishing FSI, second the more realistic case with
inclusion of a FSI phase operator.  When ignoring FSI, the wave
functions for the ejected hadrons reduce to plane waves and
$\mathcal{F}_{\text{FSI}}(\vec{r}) \equiv 1$,
$\phi_{\alpha_1}^D(\vec{p}_m) \equiv \phi_{\alpha_1}(\vec{p}_m)$.
After substituting in Eq. (\ref{matrixfinal}) the completeness
relation for Dirac spinors:
\begin{equation}
\sum_{m'_s} \left[u(\vec{p}_m,m'_s)\bar{u}(\vec{p}_m,m'_s) -
v(\vec{p}_m,m'_s)\bar{v}(\vec{p}_m,m'_s)\right] = \Id_{4\times4}\,,
\end{equation}
one obtains
\begin{align}
\left(\mathcal{M}_{fi}^{(\gamma,N\pi)}\right)_{\text{RPWIA}} = &(2\pi)^{3/2}
\sum_{m'_s}
\left(\mathcal{M}_{fi\,\mathrm{free}}^{(\gamma,N\pi)}\right)_{m_{s},m'_s}
\bar{u}(\vec{p}_m,m'_s)\phi_{\alpha_1}(\vec{p}_m)\nonumber\\
&- \mathrm{negative\,\, energy\,\, terms}\,,
\end{align}
where the RPWIA denotes the relativistic plane wave impulse
approximation.  From this last expression it is clear that even with
vanishing FSI the presence of negative-energy components makes
factorization impossible.  In what follows we neglect those terms:
\begin{equation}
\left(\mathcal{M}_{fi}^{(\gamma,N\pi)}\right)_{\text{RPWIA}} \approx
(2\pi)^{3/2} \sum_{m'_s}
\left(\mathcal{M}_{fi\,\mathrm{free}}^{(\gamma,N\pi)}\right)_{m_{s},m'_s}\bar{u}
(\vec{p}_m,m'_s)\phi_{\alpha_1}(\vec{p}_m)\,.
\end{equation}
The contraction of the Dirac spinor $\bar{u}$ with the bound nucleon wave
function $\phi_{\alpha_1}$ if negative energy components are neglected is given by
\begin{equation}
\bar{u}(\vec{p}_m,m'_s)\phi_{\alpha_1}(\vec{p}_m)=(-i)^l\sqrt{\frac{E_{N_i}
(p_m)+m_{N_i}}{2m_{N_i}}}\alpha_{n\kappa}(p_m)\chi^\dagger_{\frac{1}{2},m'_s}
\mathcal{Y}_{\kappa m}(\Omega_p,\vec{\sigma})\,,
\end{equation}
where $m_{N_i}$ is the free mass of the bound nucleon, 
$E_{N_i}(p_m)=\sqrt{m_{N_i}^2+p_m^2}$ and
\begin{equation}
\alpha_{n\kappa}(p_m)=\frac{2m_{N_i}}{E_{N_i}+m_{N_i}}g_{n\kappa}(p_m). 
\end{equation}
In this last equation $g_{n\kappa}$  is
defined as
\begin{equation}
g_{n\kappa}(p)=i \sqrt{\frac{2}{\pi}}\int_0^\infty r^2 dr
\frac{G_{n\kappa(r)}}{r}j_l(pr)\,,
\end{equation}
with $j_l(pr)$ the spherical Bessel function of the first kind.  After
squaring the matrix element and summing over the quantum number $m$ of
the bound nucleon wave function, one can use the following property of
the spin spherical harmonics $\mathcal{Y}_{\kappa m}$
\begin{equation}
\sum_m \mathcal{Y}_{\kappa m}(\Omega_p,\vec{\sigma}) \mathcal{Y}^\dagger_{\kappa
m}(\Omega_p,\vec{\sigma})=\frac{(2j+1)}{8\pi}\Id_{2\times2}\,.
\end{equation}
Finally, by using
$\chi^\dagger_{\frac{1}{2},m_s}\chi_{\frac{1}{2},m'_s}=\delta_{m_sm'_s}$,
the free pion production process can be formally decoupled from the
typical nuclear effects:
\begin{multline}
\label{matrixfull}
\overline{\sum_{fi}}
|\mathcal{M}_{fi}^{(\gamma,N\pi)}|^2=\frac{1}{2} 
\sum_{\lambda,m,m_s}|\mathcal{M}
_{fi}^{(\gamma,N\pi)}|^2 \approx
(2\pi)^3\frac{2j+1}{4\pi}\frac{E_{N_i}(p_m)+m_{N_i}}{2m_{N_i}}|\alpha_{n\kappa}
(p_m)|^2\\
\times\frac{1}{4}\sum_{\lambda,m_s,m'_s}|\left(\mathcal{M}_{fi\,\mathrm{free}}^{
  (\gamma,N\pi)}\right)_{m_s,m'_s}|^2\,.
\end{multline}
The matrix element for the pion production process on a free nucleon
is related to the cross section by
\begin{equation}
\label{matrixfree}
\frac{1}{4}\sum_{\lambda,m_s,m'_s}\mid\left(\mathcal{M}_{fi\,\mathrm{free}}^{
(\gamma,N\pi)}\right)_
{m_s,m'_s}\mid^2=\frac{4\pi
(s-m_{N_i}^2)^2}{m_{N_i}m_N}\frac{d\sigma^{\gamma\pi}}{d\mid t \mid}\,,
\end{equation}
with $s=(p_N^\mu+p_\pi^\mu)^2$ and $t=(q^\mu-p_\pi^\mu)^2$ the Mandelstam
variables of
the free process.

After substituting Eqs.~(\ref{matrixfull}) and (\ref{matrixfree}) in
Eq.  (\ref{cross}), the differential cross section for $\gamma + A
\rightarrow (A-1)+N+\pi$ in the relativistic plane wave impulse
approximation (RPWIA) reads
\begin{multline}\label{crossnoFSI}
\left(\frac{d^5\sigma}{dE_\pi d\Omega_\pi d\Omega_N}\right)_{RPWIA} \approx
\frac{M_{A-1} p_\pi p_N (s-m_{N_i}^2)^2}{4\pi m_{N_i}q
E_A}f^{-1}_{rec}\\
\times\frac{2j+1}{4\pi}\frac{\left(E_{N_i}(p_m)+m_{N_i}\right)}{2m_{N_i}}
|\alpha_ {
n\kappa }
(p_m)|^2\frac{d\sigma^{\gamma\pi}}{d\mid t\mid} \,.
\end{multline}

When FSI are included, the derivation outlined earlier is no longer
possible due to the presence of $\mathcal{F}_{\text{FSI}}(\vec{r})$ in
$\phi_\alpha^D$.  We define a distorted momentum distribution along the lines
of Ref. \cite{Vignote:2003er}
\begin{equation} \label{RMSGA}
\rho_{D}(\vec{p}_m)=
\sum_{m_s,m}|\bar{u}(\vec{p}_m,m_s)\phi^D_{\alpha_1}(\vec{p
}_m)|^2\,.
\end{equation}
When FSI and negative energy contributions to
$\phi^D_{\alpha_1}$ are neglected, Eq. (\ref{RMSGA}) reduces to
$\frac{2j+1}{4\pi}\frac{E_{N_i}(p_m)+m_{N_i}}{2m_{N_i}}|\alpha_{n\kappa}
(p_m)|^2$.  Based on this analogy, we write the differential cross section with
FSI as
\begin{equation}
\left(\frac{d^5\sigma}{dE_\pi d\Omega_\pi d\Omega_N}\right)_{D} \approx
\frac{ M_{A-1} p_\pi p_N (s-m_{N_i}^2)^2}{4\pi m_{N_i}q
E_A}f^{-1}_{rec}\rho_{D}(\vec{p}_N,\vec{p}_m)
\frac{d\sigma^{\gamma\pi}}{d\mid t
\mid} \,.\label{crossFSI}
\end{equation}

%
%----------------------------------------------------------------------
% Subsection: Electroproduction
% 
%----------------------------------------------------------------------
%
%

\subsection{Pion electroproduction}
\label{subsec:electro}
The four-momentum of the virtual photon $\gamma ^* $ is
$q^\mu(\omega,\vec{q})$ and the $z$-axis lies along $\vec{q}$.  The
incoming (scattered) electron has four-momentum
$p_e^\mu(E_e,\vec{p}_e)$ $(p_{e'}^{\mu}(E_{e'},\vec{p}_{e'}))$ and spin
$s$ $(s')$, $\theta_e$ denotes the electron scattering angle.  With
these additional notations and conventions, the differential cross
section in the lab frame reads
\begin{equation}
\frac{d^8\sigma}{d\Omega_{e'}dE_{e'}dE_\pi d\Omega_\pi
d\Omega_N}=\frac{m_e^2p_{e'}}{(2\pi)^3 p_e} \frac{M_{A-1}m_N p_\pi
p_N}{2(2\pi)^5 E_A}f^{-1}_{rec}
\overline{\sum_{fi}}
\left|
\mathcal{M}_{fi}^{(\gamma^*,N\pi)}
\right| ^2 \, ,
\end{equation}
with the recoil factor $f_{rec}$ as in Eq. (\ref{eq:recoil}).  The
invariant matrix element $\mathcal{M}_{fi}^{(e,e'N\pi)}$ can be
written as
\begin{equation}
\mathcal{M}_{fi}^{(e,e'N\pi)}=
\langle P_\pi^{\mu}, P_N^{\mu} m_s,
P_{A-1}^{\mu} J_R M_R |
j_\mu\frac{e}{Q^2}J^\mu   
| P_A^{\mu} 0^+
\rangle
\, ,
\end{equation}
with the electron current
\begin{equation}
j_\mu=\bar{u}(\vec{p}_{e'},s')\gamma_\mu u( \vec{p}_e,s)\,, 
\end{equation}
$Q^2=-q_\mu q^\mu$ and the hadron current
$J^\mu$. 
By defining an auxiliary current
\begin{equation}
a_\mu \equiv j_\mu-\frac{j_0}{\omega}q_\mu\,
\end{equation}
and using current conservation, the following identity can readily be proved:
\begin{equation}
j_\mu J^\mu = -a_iJ_i = -a_i\delta_{ij}J_j = -\sum_{ \lambda = \left( x,y,z \right)}
a_i e_i(\lambda) e_j(\lambda)J_j\,,
\end{equation}
where $\vec{e}(\lambda)$ is the unit vector along the axis
$\lambda = \left( x,y,z \right)$.  After defining the electron density
matrix
\begin{equation}
\rho_{\lambda\lambda'}=\sum_{ss'}\left[\vec{e}(\lambda)\cdot
\vec{a}\right]^\dagger\left[\vec{e}(\lambda')\cdot \vec{a}\right]
\end{equation}
and the hadronic matrix elements
\begin{equation}
w_\lambda=
\langle P_\pi^{\mu}, P_N^{\mu} m_s,
P_{A-1}^{\mu} J_R M_R |
\vec{e}(\lambda)\cdot\vec{J}
| P_A^{\mu} 0^+
\rangle
\,,
\end{equation}
we can write for the matrix element
\begin{equation}\label{element}
\sum_{ss'}\left| \mathcal{M}_{fi}^{(e,e'N\pi)} \right| ^2 = 
\frac{e^2}{Q^4}\sum_{\lambda\lambda'}\rho_{
\lambda\lambda' }
w_\lambda^\dagger w_{\lambda'}\,.
\end{equation}
With the degree of transverse polarization defined as 
\begin{equation}
\epsilon = \left(1+\frac{2q^2}{Q^2}\tan^2{\frac{\theta_e}{2}}\right)^{-1}\,,
\end{equation}
the electron density matrix becomes  \cite{Lee:1995pv}
\begin{equation}\label{matrix}
\rho_{\lambda\lambda'}=\frac{Q^2}{m_e^2}\frac{1}{1-\epsilon} \left(
\begin{array}{ccc}
\frac{1}{2}(1+\epsilon) & 0 &
-\frac{1}{2}\sqrt{2\frac{Q^2}{\omega^2}\epsilon(1+\epsilon)}\\
0 & \frac{1}{2}(1-\epsilon) & 0\\
-\frac{1}{2}\sqrt{2\frac{Q^2}{\omega^2}\epsilon(1+\epsilon)} & 0 &
\frac{Q^2}{\omega^2}\epsilon \end{array}
\right)\,.
\end{equation}
After substituting Eq. (\ref{matrix}) in Eq. (\ref{element}), one can
factor out a part containing all the variables related to the
electrons in the differential cross section:
\begin{equation}\label{factor1}
\frac{d^8\sigma}{d\Omega_{e'}dE_{e'}dE_\pi d\Omega_\pi
d\Omega_N}=\Gamma
\frac{d^5\sigma_v}{dE_\pi d\Omega_\pi
d\Omega_N} \equiv \Gamma \mathcal{C}
\overline{\sum}|M^{\gamma^*,N\pi}_{fi}|^2\,.
\end{equation}
Here, $\mathcal{M}_{fi}^{(\gamma^*,N\pi)}= \langle P_\pi^{\mu}, P_N^{\mu} m_s,
P_{A-1}^{\mu} J_R M_R |\hat{\mathcal{O}} | q^{\mu}, P_A^{\mu} 0^+
\rangle$, $\mathcal{C}=\frac{M_{A-1}m_N p_\pi
p_N}{4(2\pi)^5 E_\gamma E_A}f^{-1}_{rec}$ and
$\Gamma=\frac{\alpha}{2\pi^2}\frac{E_{e'}}{E_e}\frac{E_\gamma}{Q^2}\frac{1}{
1-\epsilon}$ is the electron flux factor, with the virtual photon equivalent
energy $E_\gamma=\frac{s-M_A^2}{2M_A}$, the fine-structure constant $\alpha$,
and $s=(q^\mu+P_A^\mu)^2$ the Mandelstam
variable of the virtual photoproduction process.  The cross
section can be cast in the following form
\begin{equation}
\frac{d^5\sigma_v}{dE_\pi d\Omega_\pi
d\Omega_N}\equiv \frac{d^5\sigma_T}{dE_\pi d\Omega_\pi
d\Omega_N} + \epsilon
\frac{d^5\sigma_L}{dE_\pi d\Omega_\pi
d\Omega_N} + \epsilon \frac{d^5\sigma_{TT}}{dE_\pi d\Omega_\pi
d\Omega_N} +
\sqrt{\epsilon(\epsilon+1)} \frac{d^5\sigma_{TL}}{dE_\pi d\Omega_\pi
d\Omega_N}\,,
\end{equation}
with
\begin{align}
\frac{d^5\sigma_T}{dE_\pi d\Omega_\pi
d\Omega_N} &= \frac{\mathcal{C}}{2} \sum_{m_s M_R}
\left[|J_x|^2+|J_y|^2\right]\,,\nonumber\\
\frac{d^5\sigma_L}{dE_\pi d\Omega_\pi
d\Omega_N} &= \mathcal{C}\frac{Q^2}{\omega^2} \sum_{m_s M_R}
|J_z|^2\,,\nonumber\\
\frac{d^5\sigma_{TT}}{dE_\pi d\Omega_\pi
d\Omega_N} &= \frac{\mathcal{C}}{2} \sum_{m_s M_R}
\left[|J_x|^2-|J_y|^2\right]\,,\nonumber\\
\frac{d^5\sigma_{TL}}{dE_\pi d\Omega_\pi
d\Omega_N} &= \frac{-\mathcal{C}}{2}\sqrt\frac{2Q^2}{\omega^2} \sum_{m_s M_R}
\left[J_x^*J_z+J_z^*J_x\right] \, . 
\end{align}
As for the photoproduction case, we wish to establish a relation
between the invariant matrix element for virtual-photon
pion-production on a nucleus $(\mathcal{M}_{fi}^{\gamma^*,N\pi})$ and on a
free nucleon $(\mathcal{M}_{fi,\text{free}}^{\gamma^*,N\pi})$. In comparison
with the
real photoproduction process, the virtual photon has an extra degree
of polarization and $Q^2 \neq 0$.  This does not alter the derivation
presented in the previous subsection and after neglecting negative
energy contributions, one arrives at
\begin{equation} \label{electrofactor}
\mathcal{M}_{fi}^{\gamma^*,N\pi}\approx
(2\pi)^{3/2}\sum_{m_{s'}}(\mathcal{M}_{fi,\text{free}}^{
\gamma^*,N\pi})_{
\lambda , m_s , m_ {s'} } \bar
{u}(\vec{p}_m,m_{s'})\phi_\alpha^D(\vec{p}_m) \, .
\end{equation}
The matrix element $\mathcal{M}_{fi,\text{free}}^{\gamma^*,N\pi}$ is related to
the free
electroproduction process by 
\begin{equation} \label{electrocrossfree}
\frac{d^5\sigma^{eN}}{dE_{e'}d\Omega_{e'}d\phi^*_\pi
d\mid t \mid}=\Gamma'\frac{m_N^2}{2(2\pi)^2(s'-m_N^2)^2} \overline { \sum }
|\mathcal{M}_{fi,\text{free}}^
{ \gamma^*,N\pi } |^2\,,
\end{equation}
where
$\Gamma'=\frac{\alpha}{2\pi^2}\frac{E_{e'}}{E_{e}}\frac{K}{Q^2}\frac{1}{
1-\epsilon}$ is the electron flux factor, with the virtual photon
equivalent energy $K=\frac{s'-m_N^2}{2m_N}$.  Further,
$s'=(p_N^\mu+p_\pi^\mu)^2$ and $t=(q^\mu-p_\pi^\mu)^2$ are the Mandelstam
variables for
the free process.  Starred variables denote center-of-mass values.

With $\rho_{\text{D}}$ defined in Eq. (\ref{RMSGA}) and by making use of
Eqs.
(\ref{electrofactor}) and (\ref{electrocrossfree}), we arrive at the factorized
form for the differential $A(e,e'N\pi)$ cross section:
\begin{equation} \label{electrocrosssection}
\left(\frac{d^8\sigma}{d\Omega_{e'}dE_{e'}dE_\pi d\Omega_\pi
d\Omega_N}\right)_{\text{D}}=\frac{\Gamma}{\Gamma'}\frac{M_{A-1}
p_Np_\pi(s'-m_N^2)^2}{2 m_N
E_\gamma
E_A}f^{-1}_{rec}\rho_{\text{D}}\frac{d^5\sigma^{eN}}{dE_{e'}d\Omega_{e'}
d\mid t
\mid d\phi^*_\pi}\,.
\end{equation}
We wish to stress that the assumptions made to arrive at this expression, are
essentially identical to those made for the real photon case discussed in the
previous subsection.
\subsection{Final-state Interactions}
\label{subsec:fsi}
The Glauber approach can be justified when the wavelength of the
outgoing hadron is sufficiently small in comparison to the typical
interaction length with the residual nucleons. In the context of
$A(e,e'p)$ reactions \cite{Lava:2004zi} it was shown that the Glauber
model represents a realistic approach to FSI for proton kinetic
energies down to about 300~MeV. This corresponds to proton de Broglie
wavelengths of the order of 1.5~fm. For pions comparable wavelengths
are reached for kinetic energies of the order of 700~MeV.

A relativistic extension of the Glauber model, dubbed the Relativistic
Multiple-Scattering Glauber Approximation (RMSGA), was introduced in
Ref.~\cite{Ryckebusch:2003fc}. In the RMSGA, the wave function for the
ejected nucleon and pion is a convolution of a relativistic plane wave
and an Glauber eikonal phase operator which accounts for FSI
mechanisms.  In Glauber theory the assumption is made that a fast
moving particle interacts through elastic or mildly inelastic
collisions with \emph{frozen} point scatterers in a target.
Scattering angles are assumed small and each of the point scatterers
adds a phase to the wave function, resulting in the following
expression for the Glauber eikonal phase:
\begin{equation}\label{glauberphase}
\widehat{\mathcal{S}}_{iN}(\vec{r},\vec{r}_2,\ldots,\vec{r}_A)=
\prod_{j=2}^A
\left[ 1-\Gamma_{iN} (\vec{b}-\vec{b}_j)\theta(z_j-z)\right] \,\;
(\textrm{with} \; i = \pi \; \textrm{or} \; N') \; .
\end{equation}
Here, $\vec{r}_j (\vec{b}_j, z_j) $ are the coordinates of the
residual nucleons and $\vec{r} (\vec{b},z) $ specifies the interaction
point with the (virtual) photon.  In Eq. (\ref{glauberphase}), the $z$
axis lies along the path of the ejected particle $i$ (the proton or pion), 
$\vec{b}$ is perpendicular to this path.  The Heaviside step
function $\theta$ guarantees that only nucleons in the forward path of the
outgoing
particle contribute to the eikonal phase.

\begin{figure}[htb]
\centering
\includegraphics[width=15cm]{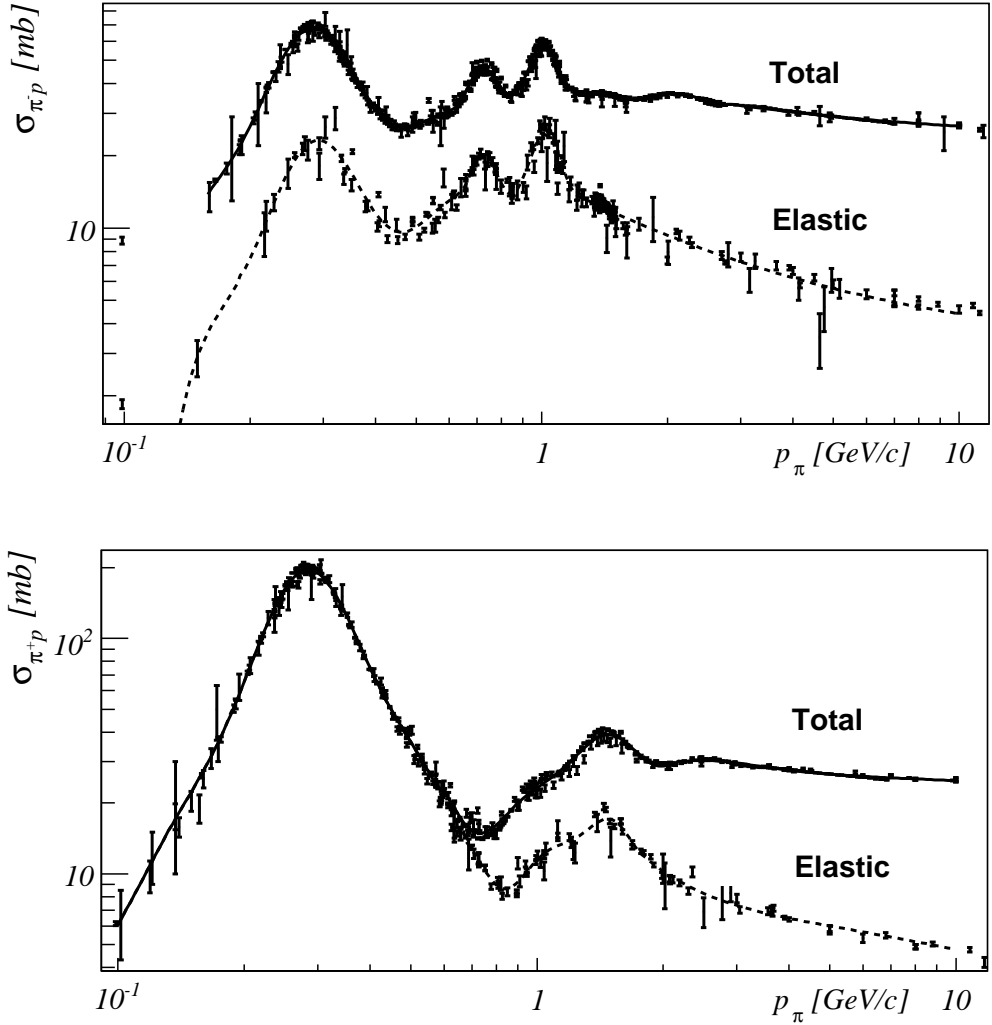}
\caption{The pion lab-momentum dependence of the data \cite{PDBook}
  and adopted fits for the total and elastic cross section
for $\pi^--p$ (upper panel) and $\pi^+-p$ (lower panel) scattering.}
\label{fig:pioncross}
\end{figure}

\begin{figure}[htb]
\centering
\includegraphics[width=15cm]{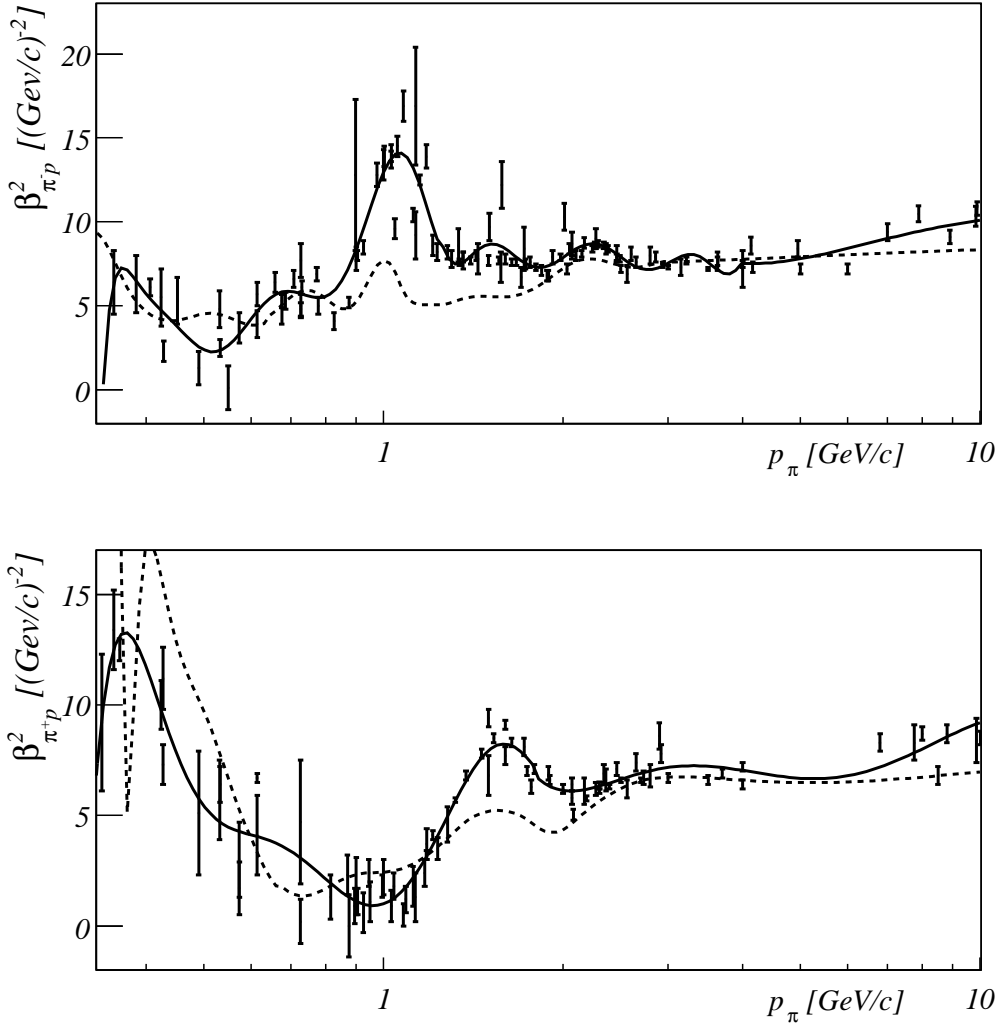}
\caption{The pion lab-momentum dependence of the data
\cite{Lasinski:1972tr} and fits for the $\beta_{p\pi}^2$ parameter for
$\pi^--p$ (upper panel) and $\pi^+-p$ (lower panel) scattering.  Full
curves are a $\chi^2$ fit to the data, whereas the dashed curves
result from Eq.~(\ref{eq:betarelation}).}
\label{fig:pionbeta}
\end{figure}

\begin{figure}[htb]
\centering
\includegraphics[width=15cm]{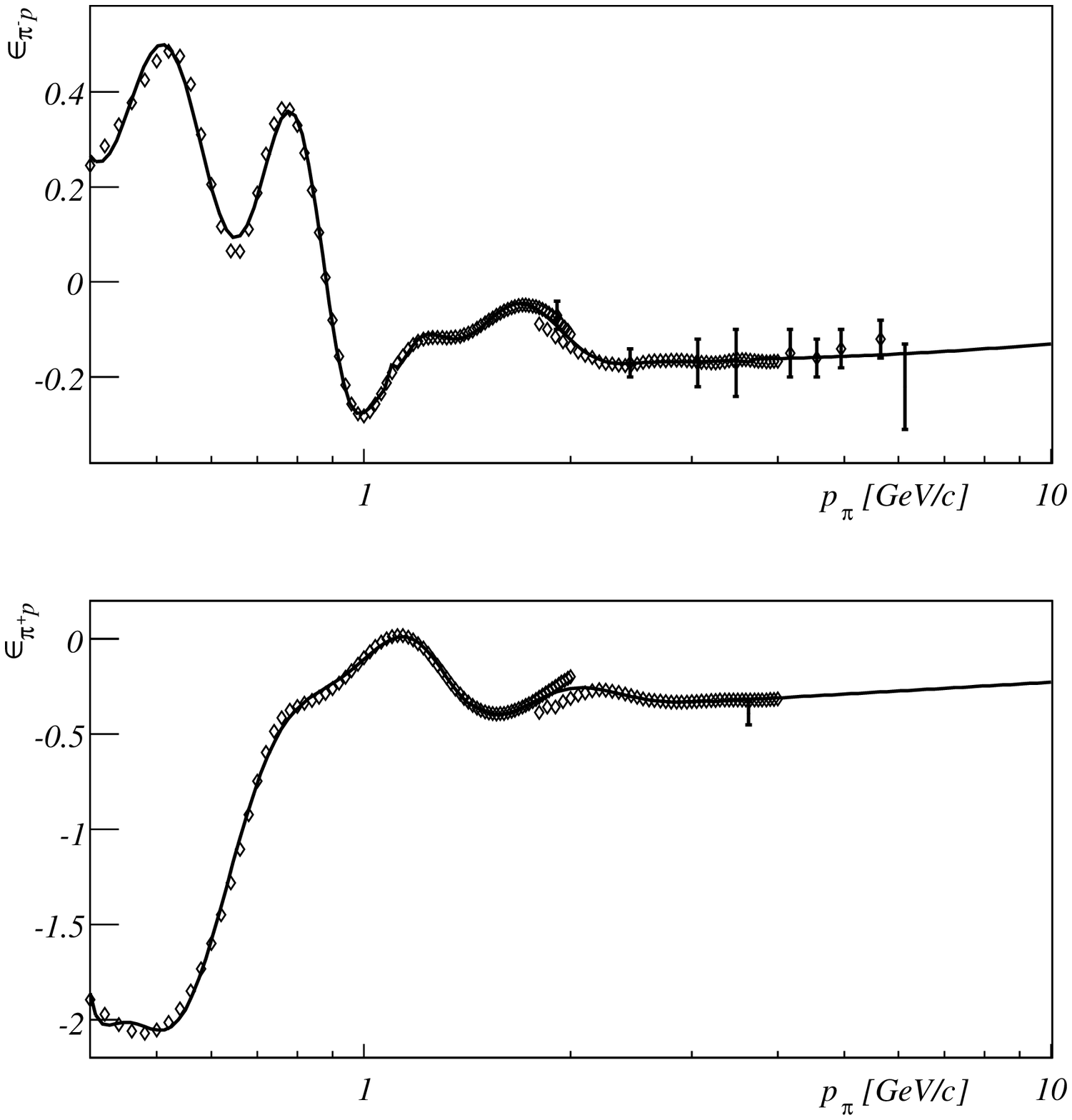}
\caption{The pion lab-momentum dependence of the ratio of the real to
imaginary part of the $\pi^-p$ (upper panel) and $\pi^+p$ (lower
panel) amplitudes.  The diamonds represent an analysis of the data by
the Georges Washington University group \cite{Arndt:2003if,Workman},
whilst the solid circles are from PDG \cite{PDBook}.  The solid line
is the fit to the data which is used in the numerical calculations.}
\label{fig:pionepsilon}
\end{figure}

Reflecting the diffractive nature of the nucleon-nucleon ($N'N$) and
pion-nucleon ($\pi N $) collisions at intermediate energies, the
profile functions $\Gamma_{N'N}$ and $\Gamma_{\pi N}$ in
Eq. (\ref{glauberphase}) are parametrized as
\begin{equation}
\label{eq:profile}
\Gamma_{iN} (\vec{b}) =
\frac{\sigma^{\text{tot}}_{iN}(1-i\epsilon_{iN})}
{4\pi\beta_{iN}^2}\exp{\left(-\frac{\vec{b}^2}{2\beta_{iN}^2}\right)}\,\;
(\textrm{with} \; i = \pi \; \textrm{or} \; N') \; .
\end{equation}
Here, the parameters $\sigma^{\text{tot}}_{iN}$ (total cross section),
$\beta_{iN}$ (slope parameter) and $\epsilon_{iN}$ (ratio of the real
to imaginary part of the scattering amplitude) depend on the momentum
of the outgoing nucleon or pion $i$. For $i=N'$, we determined the parameters
by performing a fit \cite{Ryckebusch:2003fc} to the $N' N
\longrightarrow N' N$ databases from the Particle Data Group (PDG)
\cite{PDBook}.  For the pion, $\sigma^{\text{tot}}_{\pi N}$ was fitted
to data collected by PDG \cite{PDBook}.  The analysis of the slope
parameter in Ref. \cite{Lasinski:1972tr} was used for the $\beta_{\pi
N}$ fits.  Fits provided by SAID \cite{Arndt:2003if,Workman} and data
from PDG \cite{PDBook} were used in constructing the fits for
$\epsilon_{\pi N}$.  The fits for $\sigma^{\text{tot}}_{iN},
\beta_{iN}$ and $\epsilon_{iN}$ of Figs.
\ref{fig:pioncross},\ref{fig:pionbeta} and \ref{fig:pionepsilon} are
the result of a $\chi^2$ minimization of the data against a a $n$-th
degree polynomial (with $n \leq 10$).  An alternative way of
determining $\beta_{\pi N}$, is via the relation
\begin{equation}\label{eq:betarelation}
 \beta^2_{\pi N}= \frac{(\sigma^{\text{tot}}_{\pi N})^2
(1+\epsilon^2_{\pi
N})} {16\pi\sigma^{\text{el}}_{\pi N}}\,,
\end{equation}
with $\sigma^{\text{el}}_{\pi N}$ the elastic cross section. 
Fits for $\sigma^{\text{el}}_{\pi N}$ to data
from PDG \cite{PDBook} are also presented in Fig.~\ref{fig:pioncross}.
The two sets for the $\beta_{\pi N}$ parameter in  Fig. \ref{fig:pionbeta} do
not produce
significantly different results for the numerical calculations
presented here. We use the
$\chi^2$ fit for $\beta_{\pi N}$ in all calculations presented in this paper.

The Glauber operator of Eq. (\ref{glauberphase}) is an $A$-body
operator. As a consequence, it requires integrations over all
spectator nucleon coordinates in Eq.~(\ref{glauberdef}), which is
computationally very demanding, in particular for heavy target nuclei.
In $\gamma^{(*)}+A\rightarrow (A-1)+N+\pi$ calculations, a product
of two Glauber phases is involved and the cylindrical symmetry of the
individual phases is lost. A Romberg algorithm is used to perform the
integrations over the spatial coordinates in Eq.~(\ref{glauberdef}).

For nucleons with a kinetic energy lower than about 300~MeV, the
approximations underlying the Glauber formalism are no longer
applicable, and an alternative method to model FSI is required. Under
those circumstances our framework provides the flexibility to adopt
the Relativistic Optical Model Eikonal Approximation (ROMEA)
\cite{VanOvermeire:2006dr}. In the ROMEA approach, the wave function
of a nucleon with energy $E = \sqrt{p_N^2+m_N^2}$ after scattering in
a scalar ($V_s(r)$) and vector ($V_v(r)$) spherical potential has the
following form:
\begin{equation}
\label{scatwave}
\psi^{(+)}_{\vec{p}_N,m_s} ( \vec{r} ) = 
\sqrt{\frac{E+m_n}{2m_N}}
\left[
\begin{array}{c}
1 \\ \frac{1}{E+m_N+V_s(r)-V_v(r)}\vec{\sigma}\cdot \hat{\vec{p}}
\end{array}
\right] 
e^{i\vec{p}_N\cdot\vec{r}}e^{i\hat{\mathcal{S}}_{N'N} 
(\vec{r})}\chi_{\frac{1}{2}
m_s}\,,
\end{equation}
with the eikonal phase determined by 
\begin{equation}
\label{eikonal}
i \hat{\mathcal{S}}_{N'N}(\vec{b},z) = -i \frac{m_N}{K}\int_{-\infty}^z dz'
\left[V_c(\vec{b},z')+V_{\text{so}}(\vec{b},z')\left[\vec{\sigma}\cdot(\vec{b}
\times \vec{K})-iKz'\right]\right]\,.
\end{equation}
In this last equation, $\vec{K}=\frac{1}{2}(\vec{k}_i+\vec{k}_f)$ is
the average of the initial and final momentum of the scattering
particle.  In the small angle approximation, $\vec{K}\approx
\vec{p}_N$ and points along the z-axis.  The central and spin-orbit
potentials $V_c$ and $V_{so}$ are functions of $V_s$ and $V_v$ and
their derivatives \cite{VanOvermeire:2006dr}.

Additional approximations were used in the implementation of
optical-potential FSI in this ROMEA model.  The dynamical enhancement
of the lower components of the scattering wave function
(\ref{scatwave}) is ignored as at low momenta the lower components are
small compared to the upper components due to $\hat{\vec{p}}$ and at
higher momenta $(V_s-V_v)$ is small in comparison to $(E+m_N)$.  The
operator $\hat{\vec{p}}$ was also substituted by the asymptotic value
$\vec{p}_N$.  Finally, as collisions were assumed spin-independent in
(\ref{eq:approx}), the spin-orbit potential $V_{so}$ in
Eq. (\ref{eikonal}) is neglected.  This yields the following phase
factor entering in Eq. (\ref{glauberdef}):
\begin{equation} \label{eq:romeaphase}
\hat{\mathcal{S}}_{N'N}(\vec{r})=e^{-i\frac{m_N}{p_N}\int_{z_N}^{+\infty}dz
V_c(\vec{b}_{p_N},z)}\,.
\end{equation}

In contrast to the Glauber eikonal phase, the optical potential
eikonal phase of Eq. (\ref{eq:romeaphase}) depends solely on the
coordinate $\vec{r}$ which defines the interaction point. As a
consequence, it can be taken out of all the integrations in
Eq. (\ref{glauberdef}) and the cylindrical symmetry of the pion
Glauber eikonal factor is retained, hereby considerably reducing the
cost of computing the total FSI factor $\mathcal{F}_{\text{FSI}}$. For
the numerical evaluation of the ROMEA phase factor, we made use of the
optical potential of van Oers \emph{et al.}  \cite{vanOers:1981mr} for
$^4\text{He}$ and the global $(S-V)$ parametrization of Cooper
\emph{et al.} \cite{Cooper:1993nx} for heavier nuclei. 

%
% Section on CT and SRC ...
%
%
\subsection{Color transparency and short-range correlations}
\label{subsec:ct}
\begin{figure}[htb]
\centering
\includegraphics[width=8cm]{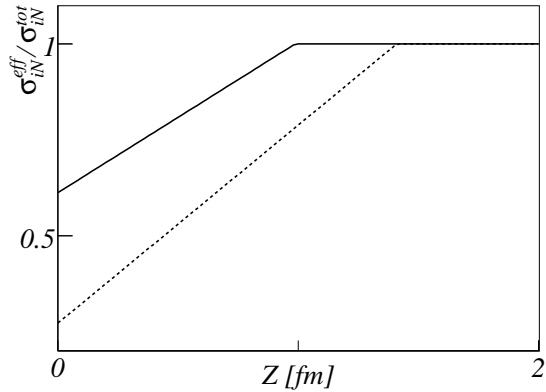}
\caption{Comparison of the CT effect on the total effective cross
section $\sigma_{iN}^{\text{eff}}$ for nucleon-nucleon (full) and
pion-nucleon (dashed) interactions.  We consider the situation whereby
the ejectile possesses a lab-momentum of 2.5~GeV/c. For the hard-scale
parameter we adopt $\mathcal{H}=1.8~\text{(GeV/c)}^2$.}
\label{fig:CT}
\end{figure}

We implement color transparency effects in the usual fashion by
replacing the total cross sections $\sigma^{\text{tot}}_{iN}$ in the
profile functions of Eq.~(\ref{eq:profile}) with effective ones
\cite{Farrar:1988me}. The latter induce some reduced pion-nucleon and
nucleon-nucleon interaction over a typical length scale $l_h$
corresponding with the hadron formation length ($i = \pi \;
\textrm{or} \; N'$)
\begin{equation}
\frac { \sigma^{\text{eff}}_{iN} }  
{ \sigma^{\text{tot}}_{iN} } =  \biggl\{ \biggl[
 \frac{\mathcal{Z}}{l_h} + \frac{<n^2 k_t^2>}{\mathcal{H}} 
\left( 1-\frac{\mathcal{Z}}{l_h} \right) \biggr] \theta(l_h-\mathcal{Z})  + 
\theta(\mathcal{Z}-l_h) \biggr\} \,  \; .
\label{eq:diffusion}
\end{equation}
Here, $n$ is the number of elementary fields (2 for the pion, 3 for the
nucleon), $k_t = 0.350~\text{GeV/c}$ is the average transverse
momentum of a quark inside a hadron, $\mathcal{Z}$ is the distance
from the interaction point and $l_h \simeq 2p/\Delta M^2$ is the
hadronic expansion length, with $p$ the momentum of the final hadron
and $\Delta M^2$ the mass squared difference between the intermediate
prehadron and the final hadron state.  We adopt the values $\Delta M^2
= 1~ \text{GeV}^2$ for the proton and $\Delta M^2 = 0.7~\text{GeV}^2$
for the pion.  $\mathcal{H}$ is the hard scale parameter that governs
the CT effect.  It equals the momentum transfer $t=(q^\mu-p_\pi^\mu)^2$ (pion
CT) or
$u=(q^\mu-p_N^\mu)^2$ (nucleon CT) for pion
photoproduction and $Q^2$ for pion electroproduction.
Fig.~\ref{fig:CT} illustrates the predicted difference of the CT
effect on the pion-nucleon and nucleon-nucleon effective interaction.
Reflecting its mesonic nature, the pion has a longer formation length
and during its formation its interaction cross section with the
residual nucleons is more strongly reduced than for a nucleon.

We now proceed with introducing a method which allows us to implement
the effect of SRC in the relativistic Glauber calculations.  The
proposed method adopts the thickness approximation as a starting
point.  In the thickness approximation, the density $\left|
\rho_{\alpha_i}(\vec{r}_i)\right|^2$ of the individual nucleons in
Eq.~(\ref{glauberdef}) is replaced by an averaged density
$\rho_A ^{[1]}(\vec{r})$ defined as
\begin{equation}
\rho_A^{[1]}(\vec{r})=A\int d\vec{r}_2 \ldots \int d\vec{r}_A
\left( \Psi_A^{\text{g.s.}}(\vec{r},\vec{r}_2,\ldots,\vec{r}_A)\right)^\dagger
\Psi_A^{\text{g.s.}}(\vec{r},\vec{r}_2,\ldots,\vec{r}_A)\,.
\end{equation}

In terms of $\rho_A ^{[1]} (\vec{r})$ the FSI factor of
Eq.~(\ref{glauberdef}) can be approximated by
\begin{equation}
\mathcal{F}_{\text{FSI}}^{\text{thick}} \left( \vec{r} \right) 
=  \frac{1}{A^{A-1}} \int
d\vec{r}_2 \ldots \int d\vec{r}_A
\rho^{[1]}_A(\vec{r}_2)\rho^{[1]}_A(\vec{r}_3)\ldots
\rho^{[1]}_A(\vec{r}_A)\hat{\mathcal{S}}_{\pi
N}(\vec{r};\vec{r}_2,\ldots,\vec{r}_A)\hat{\mathcal{S}}_{N'
N}(\vec{r};\vec{r}_2,\ldots,\vec{r}_A)
\end{equation}
In combination with the operators of Eq.~(\ref{glauberphase}) the
expression can be further simplified to
\begin{multline}
\label{fsithick}
\mathcal{F}_{\text{FSI}}^{\text{thick}} \left( \vec{r} \right) = 
 \left(\int d\vec{r}_2
\frac{\rho^{[1]}_A(\vec{r}_2)}{A} 
\left[ 1-\Gamma_{N'p}(\vec{b}_{N'}-\vec{b}_{N'2}
)\theta(z_{N'2}-z_{N'}) \right] \right.\\
\left.\left[1-\Gamma_{\pi p}(\vec{b}_{\pi}-\vec{b}_{\pi
2})\theta(z_{\pi 2}-z_\pi)\right]\right)^{Z-\frac{\tau_z+1}{2}}\\
\times\left(\int d\vec{r}_3
\frac{\rho^{[1]}_A(\vec{r}_3)}{A}\left[1-\Gamma_{N'n}(\vec{b}_{N'} - 
\vec{b}_{N'3}
)\theta(z_{N'3}-z_{N'})\right]\right.\\
\left.\left[1-\Gamma_{\pi n}(\vec{b}_{\pi}-\vec{b}_{\pi
3})\theta(z_{\pi
3}-z_\pi)\right]\right)^{N+\frac{\tau_z-1}{2}}\,,
\end{multline}
where $\tau_z$ is the isospin (1 for protons and -1 for neutrons) of
the nucleon on which the initial absorption took place.  The $z_{N'}$
($z_\pi$) axis lies along the ejected nucleon (pion).  The above
expression is derived within the context of the IPM.  It is clear that
the nucleus has a fluid nature and that the IPM can only be considered
as a first-order approximation.  In computing the FSI effects by means
of the Eq. (\ref{fsithick}) one fails to give proper attention to one
important piece of information: namely that one considers the density
distribution of nucleons given that there is one present at the
photo-interaction point $\vec{r}$.

\begin{figure}[htb]
\centering
\includegraphics[width=14cm]{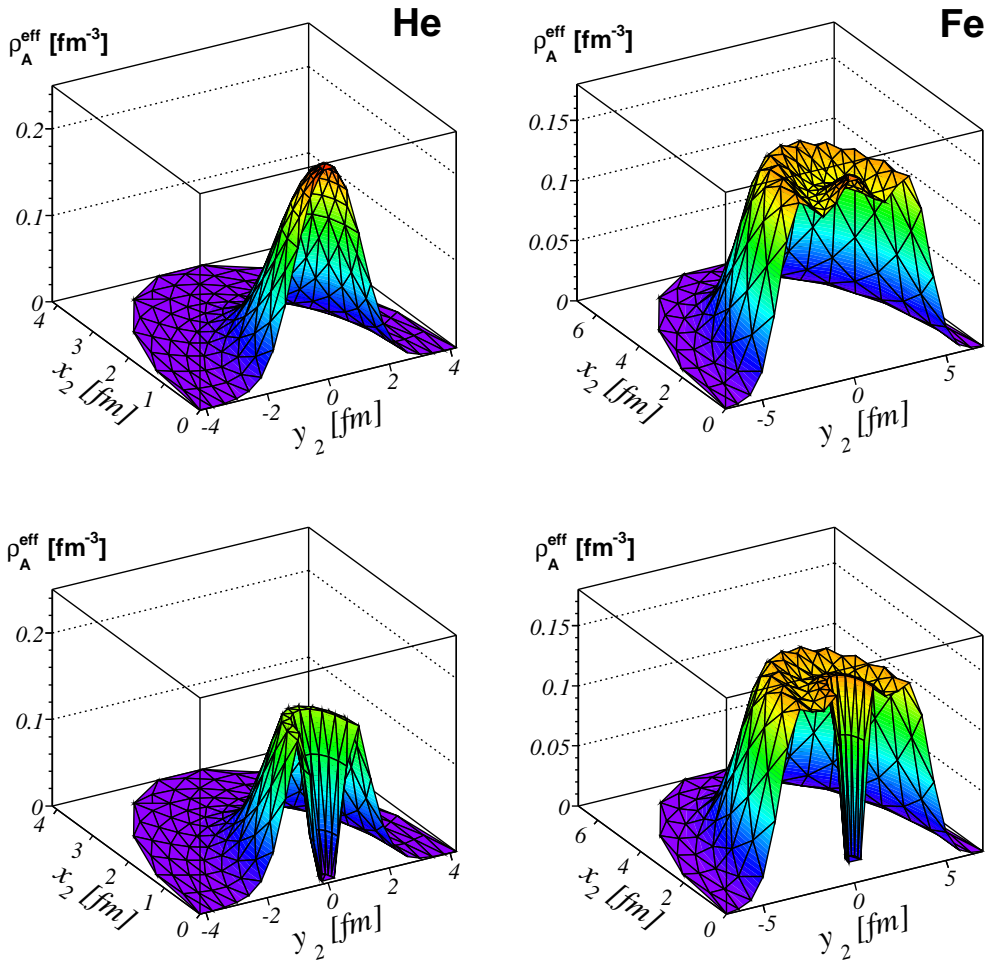}
\caption{(Color online) The effective nuclear density
$\rho^{\text{eff}}_A(\vec{r}_2, \vec{r})$ at $z_2=0$ for He (left) and
Fe (right) before (upper) and after (lower panel) the inclusion of SRC effects.
The effective nuclear densities here refer to the situation whereby the
(virtual) photon is absorbed at the origin $(x=0,y=0,z=0)$.}
\label{fig:SRC}
\end{figure}

The two-body density $\rho^{[2]}_A(\vec{r}_1,\vec{r}_2)$ is related to
the probability to find a nucleon at position $\vec{r}_2$ given that
there is one at a position $\vec{r}_1$.  We adopt the following
normalization convention for $\rho^{[2]}_A$
\begin{equation}\label{eq:normrho2}
 \int d\vec{r}_1 \int d \vec{r}_2 \rho^{[2]}_A(\vec{r}_1,\vec{r}_2)=A(A-1)\,.
\end{equation}
 In the IPM on has
$\left(\rho^{[2]}_A(\vec{r}_1,\vec{r}_2)\right)_{\text{IPM}}\equiv
\frac{A-1}{A}\rho^{[1]}_A(\vec{r}_1)\rho^{[1]}_A(\vec{r}_2)$.  The
nucleus has a granular structure as the nucleons have a finite size.
This gives rise to strong nucleon-nucleon repulsions at short internucleon
distances which
reflect themselves in short-range correlations (SRC) at the nuclear
scale.  One can correct $\left(\rho^{[2]}_A(\vec{r}_1,\vec{r}_2)\right)$ for the
presence of the SRC by adopting the following functional form 
\cite{Frankel:1992er}
\begin{equation}\label{eq:rho2SRC}
 \rho^{[2]}_A(\vec{r}_1,\vec{r}_2) \equiv
\gamma(\vec{r}_1)\left(\rho^{[2]}_A(\vec{r}_1,\vec{r}_2)\right)_{\text{IPM}}
\gamma(\vec{r}_2) g(r_{12}) =
\frac{A-1}{A}\gamma(\vec{r}_1)\rho^{[1]}_A(\vec{r}_1)\rho^{[1]}_A(\vec{r}
_2)\gamma(\vec{r}_2) g(r_{12})\,,
\end{equation}
with $g(r_{12})$ the so-called Jastrow correlation function and
$\gamma(\vec{r})$ a function which imposes the normalization condition of Eq.
(\ref{eq:normrho2}) on $\rho^{[2]}_A(\vec{r}_1,\vec{r}_2)$.  The function
$\gamma(\vec{r})$ is a solution to the following integral equation
\begin{equation}
 \gamma(\vec{r}_1)\int d\vec{r}_2 \rho^{[1]}_A(\vec{r}
_2) g(r_{12}) \gamma(\vec{r}_2)=A\,,
\end{equation}
which can be solved numerically.  The Glauber phase factor of Eq.
(\ref{fsithick}) can now be corrected for SRC through the following
substitution
\begin{equation}\label{eq:denssubst}
 \rho^{[1]}_A(\vec{r}_2) \rightarrow
\frac{A}{A-1}\frac{\rho^{[2]}_A(\vec{r}_2,\vec{r})}{\rho^{[1]}_A(\vec{r})}
=\gamma(\vec{r}_2) \rho^{[1]}_A(\vec{r}_2) \gamma(\vec{r})
g(|\vec{r}_2-\vec{r}|) \equiv \rho^{\text{eff}}_A(\vec{r}_2, \vec{r}) \; ,
\end{equation}
whereby $\rho^{[2]}_A(\vec{r}_2,\vec{r})$ adopts the expression
(\ref{eq:rho2SRC}).  These manipulations amount to the following final
expression for the Glauber FSI factor including SRC:
\begin{multline}
 \mathcal{F}_{\text{FSI}}^{\text{SRC}} \left( \vec{r} \right) =  
\Biggl( 
\int d\vec{r}_2
\frac{\gamma(\vec{r})_2 \rho^{[1]}_A(\vec{r}_2) \gamma(\vec{r})
g(|\vec{r}_2-\vec{r}|)}{A}\left[1-\Gamma_{N'p}(\vec{b}_{
N'}-\vec{b}_{N'2}
)\theta(z_{N'2}-z_{N'})\right] \\
\times 
\left[1-\Gamma_{\pi p}(\vec{b}_{\pi}-\vec{b}_{\pi
2})\theta(z_{\pi 2}-z_\pi)\right]
\Biggr) ^ {Z-\frac{\tau_z+1}{2}}\\
\times \Biggl( 
\int d\vec{r}_3
\frac{\gamma(\vec{r})_2 \rho^{[1]}_A(\vec{r}_2) \gamma(\vec{r})
g(|\vec{r}_2-\vec{r}|)}{A}\left[1-\Gamma_{N'n}(\vec{b}_{N'}-\vec{b}_{N'3}
)\theta(z_{N'3}-z_{N'})\right] \\
\times \left[1-\Gamma_{\pi n}(\vec{b}_{\pi}-\vec{b}_{\pi
3})\theta(z_{\pi
3}-z_\pi)\right] 
\Biggr)^{N+\frac{\tau_z-1}{2}}\,. 
\end{multline}
The effective density of Eq. (\ref{eq:denssubst}) accounts for the
fact that the motion of each nucleon does depend on the presence of
the other ones.  In Fig. \ref{fig:SRC} we display the effective
nuclear density as it would be observed by a nucleon or a pion created
after photoabsorption on a nucleon at the center of the nucleus.  The
figure shows the density for Fe as computed in the IPM
($\rho^{[1]}_A(x,y,z\equiv 0)$) and with the expression based on the
substitution of Eq.~(\ref{eq:denssubst})  
\begin{displaymath}
\gamma(x,y,z\equiv 0) \rho^{[1]}_A(x,y,z\equiv 0) 
\gamma(x\equiv 0,y\equiv 0,z\equiv 0) g(|\vec{r}|) \; .
\end{displaymath}
In Fig.~\ref{fig:SRC} and all forthcoming numerical calculations we
use a correlation function $g(|\vec{r}|) $ from
Ref.~\cite{Gearhart}. It is characterized by a (Gaussian) hard core of
about 0.8~fm and a second bump which extends to internucleon distances
$r$ of about 2~fm and reaches its maximum for $r_{12} \approx
1.3$~fm. This correlation function provided a fair description of the
SRC contributions to $^{12}$C$(e,e'pp)$ \cite{Blomqvist:1998gq} and
$^{16}$O$(e,e'pp)$ \cite{Ryckebusch:2003tu}.  It is clear that the SRC
lead to a local reduction - with size of the nucleon radius - of the
density around the nucleon struck by the (virtual) photon.  In order
to preserve the proper normalization, this reduction amounts to some
enhanced density at distances of about twice the nucleon radius.  With
regard to the intranuclear attenuation, the reduction of the density
in the proximity of the struck nucleon will result in some enhanced
transparency close to the photo-interaction point $\vec{r}$.  The enhanced
density
at positions of about
twice the nucleon radius from the struck nucleon, can be expected to have
the opposite effect.
%

%
% Section with the Numerical results
%
%

\section{Numerical Results}
\label{sec:results}
\subsection{The FSI factor}

In this subsection we present a selected number of results of the
numerical calculations of the RMSGA FSI factor of
Eq.~(\ref{glauberdef}).  We consider the
$^{12}\text{C}(\gamma,p\pi^-)$ reaction in a reference frame with the
$z$ axis along the momentum $\vec{p}_N$ of the ejected nucleon and the
$y$ axis along $\vec{p}_N \times \vec{p}_\pi$.  The coordinate
$\vec{r}$ denotes the interaction point with the external photon.  The
FSI factor is plotted versus the spherical coordinates in this frame.

\begin{figure}[htb]
\centering
\includegraphics[width=15cm]{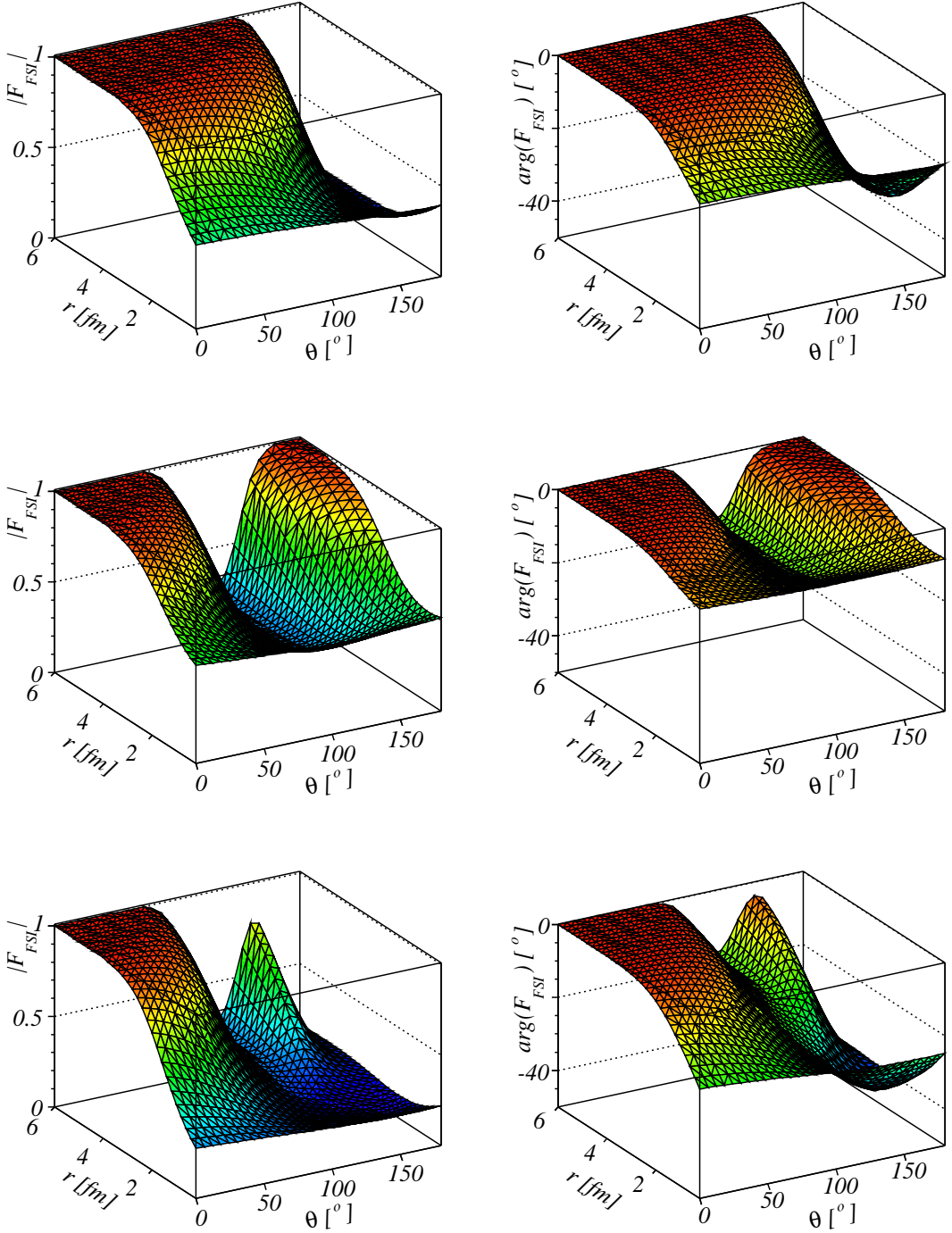}
\caption{(Color online) Radial and polar-angle dependence of the norm
(left) and phase (right) of the FSI factor $\mathcal{F}_{\text{FSI}}$
in the scattering plane ($\phi=0^o$) for the
$^{12}\text{C}(\gamma,p\pi^-)$ reaction from the $1s_{1/2}$ level.
For the upper (middle) panels, solely the FSI effects on the ejected
proton (pion) are considered. The lower panels include the net effect
of both the pion and nucleon FSI effect.  The results are obtained for
$p_N=2638$ MeV, $p_\pi = 2291$ MeV, $\theta_{N\pi} = -65.19^o$.}
\label{fig:glaubercompare}
\end{figure}

In Fig.~\ref{fig:glaubercompare}, we present the calculated norm and
phase of the FSI factor in the scattering plane ($\phi=0$) for $p_N
\approx 2.6$ GeV and $p_\pi \approx 2.3$ GeV, which are conditions for
which Jefferson Lab collected data. We present the FSI factor for the
proton and the pion separately as well as the combined effect when the
two are detected in coincidence.

When looking at the $\theta$ dependence, it becomes clear from
Fig.~\ref{fig:glaubercompare} that the norm is smallest in the
direction opposite the momentum of the particle (being $180^o$ for the
nucleon and $180^o-\theta_{N\pi}$ for the pion).  For these directions
and large $r$, the nucleon or pion is created close to the surface of
the nucleus on the opposite side of its asymptotic direction and has
to travel through a thick layer of nuclear medium before it reaches a
free status.  As for the $r$ dependence, we see for the nucleon a
reduction of the FSI effects for rising $r$ at angles in the
neighborhood of $\theta=0^o$, respectively an increment for rising $r$
at $\theta=180^o$.  This is again due to the fact that the outgoing
nucleon traverses less, respectively more nuclear matter on its way
out of the nucleus. The same observations apply for the pion, albeit
at the angles $\theta_{N\pi}$ and $180^o-\theta_{N\pi}$.  The total
FSI factor combines the intranuclear attenuation effects on the
nucleon and pion. Hence, the norm shows the largest reduction at
$\theta$ around $180^o$ and $180^o-\theta_{N\pi}$.  The phase of the
FSI factor exhibits similar behavior, with the largest phase shifts
occurring at the discussed angles.

\begin{figure}[htb]
\centering
\includegraphics[width=7cm]{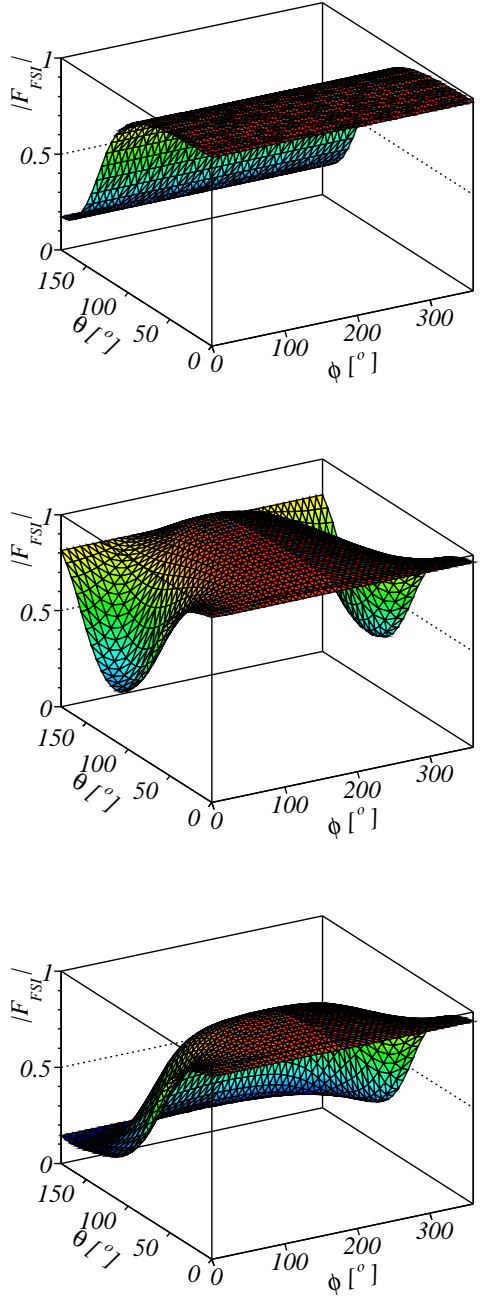}
\caption{(Color online) Polar- and azimuthal-angle dependence of the
norm of the FSI factor $\mathcal{F}_{\text{FSI}}$ at a distance $r=3$
fm from the center of the nucleus for the
$^{12}\text{C}(\gamma,p\pi^-)$ reaction from the $1s_{1/2}$ level.
Separate contributions from the nucleon (upper panel) and the pion
(middle panel), as well as their combined effect (bottom panel) are
shown. Kinematics as in Fig.~\ref{fig:glaubercompare}.}
\label{fig:glaubercompare2}
\end{figure}

Fig. \ref{fig:glaubercompare2} teaches us a couple things about the
$\phi$ dependence of the FSI factor.  As the outgoing nucleon lies
along the $z$ axis there is no dependence on the azimuthal angle
because of the cylindrical symmetry.  Again, we can see that the
absorption is largest when large amounts of nuclear matter need to be
traversed (i.e. large $\theta$).  Looking at the pion we see the
largest attenuation occurs in the upper hemisphere ($\cos{\phi}\geq
0$) as a pion that is created in this region has to traverse the inner
core of the nucleus.  The combined effect of the pion and nucleon
contributions is contained in the bottom panel.  As the reaction takes
place in the $xz$ plane, the total FSI factor retains the following
symmetry: $\mathcal{F}_{\text{FSI}}(r,\theta,\phi) =
\mathcal{F}_{\text{FSI}}(r,\theta,2\pi-\phi)$.

%
% Numerical results for pion photoproduction
%
%

\subsection{Pion photoproduction}
The experiment E94-104 at Jefferson Lab extracted nuclear
transparencies for $\gamma + ^4\text{He} \rightarrow p + \pi^- +
^3\text{He}$.  The measurements were performed for photon energies $
1.6 \le q \le 4.2 $~GeV and for center-of-mass angles
$\theta_{\mathrm{c.m.}}=70^o$ and $90^o$. In total, the nuclear
transparencies were measured for eight kinematical settings. In a
proposal for a follow-up experiment, seven additional kinematics are
suggested for measurements at higher photon energies and
$\theta_{\mathrm{c.m.}}=90^o$ \cite{Dutta}.  We have performed calculatios for
the completed and planned experiments. 
Table~\ref{tab:kinphoto}
provides a list of the kinematics.

\begin{table}[htpb]
    \begin{center}
        \begin{tabular}{lrrrrr}
            \hline
            \hline
	    $q$ & $\theta_{\text{c.m.}}$ & $p_N$ & $\theta_N$ & $p_\pi$ &
$\theta_\pi$\\
	    \hline
	    1648 & $70^o$ & 989 & $47.39^o$ & 1238 & $-36.02^o$ \\
	    1648 & $90^o$ & 1277 & $37.37^o$ & 1015 & $-47.73^o$ \\
	    2486 & $70^o$ & 1322 & $44.37^o$ & 1794 & $-31.02^o$ \\
	    2486 & $90^o$ & 1740 & $34.45^o$ & 1438 & $-43.18^o$ \\
	    3324 & $70^o$ & 1642 & $41.74^o$ & 2363 & $-27.56^o$ \\
	    3324 & $90^o$ & 2195 & $32.01^o$ & 1866 & $-38.57^o$ \\
	    4157 & $70^o$ & 1949 & $39.51^o$ & 2929 & $-25.05^o$ \\
	    4157 & $90^o$ & 2638 & $30.01^o$ & 2291 & $-35.18^o$ \\
	    4327 & $70^o$ & 2011 & $39.1^o$ & 3044 & $-24.6^o$\\
	    4327 & $90^o$ & 2727 & $29.6^o$ & 2377 & $-34.6^o$\\
 	    5160 & $70^o$ & 2307 & $37.3^o$ & 3606 & $-22.8^o$\\
 	    5160 & $90^o$ & 3161 & $28.0^o$ & 2797 & $-32.1^o$\\
 	    6059 & $70^o$ & 2622 & $35.6^o$ & 4211 & $-21.2^o$\\
 	    6059 & $90^o$ & 3625 & $26.6^o$ & 3250 & $-29.9^o$\\
 	    7025 & $70^o$ & 2956 & $33.9^o$ & 4861 & $-19.8^o$\\
 	    7025 & $90^o$ & 4120 & $25.2^o$ & 3735 & $-28.0^o$\\
 	    8057 & $70^o$ & 3309 & $32.4^o$ & 5555 & $-18.6^o$\\
 	    8057 & $90^o$ & 4646 & $24.0^o$ & 4253 & $-26.3^o$\\
 	    9156 & $70^o$ & 3683 & $31.0^o$ & 6294 & $-17.6^o$\\
 	    9156 & $90^o$ & 5204 & $22.8^o$ & 4805 & $-24.8^o$\\
 	    10322 & $70^o$ & 4077 & $29.7^o$ & 7077 & $-16.6^o$\\
 	    10322 & $90^o$ & 5794 & $21.8^o$ & 5389 & $-23.5^o$\\
	    \hline
            \hline
        \end{tabular}
    \end{center}
\caption{Central values for the photon energy (MeV), proton momentum $p_N$
(MeV), proton
angle $\theta_N$, pion momentum $p_\pi$ (MeV) and pion angle $\theta_\pi$ for
$\theta_{\mathrm{c.m.}}=70^o, 90^o$. 
Angles are measured relative to the incoming photon momentum.} 
\label{tab:kinphoto}
\end{table}

We aim at performing calculations which match the kinematic conditions of
the experiment as closely as possible. We use the following
definition for the transparency:
\begin{equation}\label{eq:phototransp}
T=\frac{ \sum_{\alpha} \int  dq Y(q) \int d\vec{p}_m  
  \left(\frac{d^5\sigma}{dE_{\pi_i} d\Omega_{\pi_i}
  d\Omega_{N_i}}\right)_{\text{RMSGA}}} {\sum_{\alpha} \int  dq Y(q) \int d\vec{p}_m  
  \left(\frac{d^5\sigma}{dE_{\pi_i} d\Omega_{\pi_i}
  d\Omega_{N_i}}\right)_{\text{RPWIA}}}\;\;.
\end{equation}
The integrations $\int  dq \int d\vec{p}_m$
in Eq. (\ref{eq:phototransp}) were evaluated with a random integration algorithm.  
To this end, random events within the photon beam energy range,
detector acceptances and applied cuts for each data point
were generated for the calculation of the transparency until
convergence of the order of 5\% was reached.  Typically, this involves about a
thousand events for each data point.
In Eq. (\ref{eq:phototransp}), $\sum _{\alpha} $ extends over
all occupied single-particle states in the target nucleus.  All cross
sections are computed in the lab frame.  $Y(q)$ provides the
weight factor for the generated events.  It includes the yield of the reconstructed experimental
photon beam
spectrum \cite{Dutta:2003mk} for the photon energy of the generated event.  We
assume that the elementary $\gamma +n \rightarrow \pi ^- + p$ cross section
$\frac { d \sigma
^{\gamma \pi}} {d \mid t \mid}$ in Eqs.~(\ref{crossnoFSI}) and (\ref{crossFSI})
remains
constant over the kinematical ranges $\int  dq \int d\vec{p}_m$ which define a
particular data
point. With this assumption the cross section $ \frac { d \sigma
^{\gamma \pi}} {d \mid t \mid} $ cancels out of the ratio
(\ref{eq:phototransp}).  For all kinematic conditions of
Table~\ref{tab:kinphoto}, the pion and nucleon momenta are sufficiently high
for the RMSGA method to be a valid approach for describing the FSI mechanism.

\begin{figure}[htb]
\centering
\includegraphics[width=7cm]{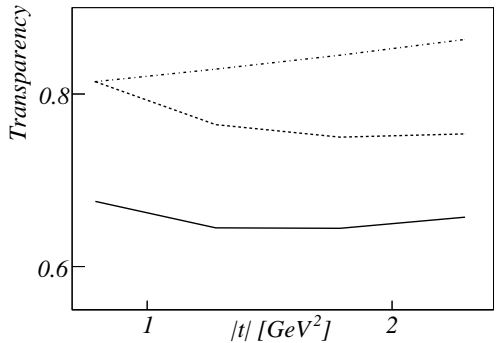}
\caption{Contributions of the pion (dashed-dotted) and nucleon
(dashed) to the total nuclear transparency (full) extracted from
$^4\text{He}(\gamma,p\pi^-)$ versus $\mid t \mid$ at
$\theta_{\text{c.m.}}=70^o$.  All calculations include CT.}
\label{fig:photocompare70}
\end{figure}

For a discussion of the computed results compared to the experimental
data and a semi-classical model we refer the reader to
Ref.~\cite{Cosyn:2006vm}.  In Fig.~\ref{fig:photocompare70} the
separated transparencies for the outgoing proton and pion are
displayed next to the full result.  It is clear from this figure that
the rise of the transparency at low $|t|$ can be attributed to the proton
contribution.  This rise can be attributed to the local minimum in the total
nucleon-nucleon cross section for nucleon momenta of about 1 GeV

\begin{figure}[htb]
\centering
\includegraphics[width=14cm]{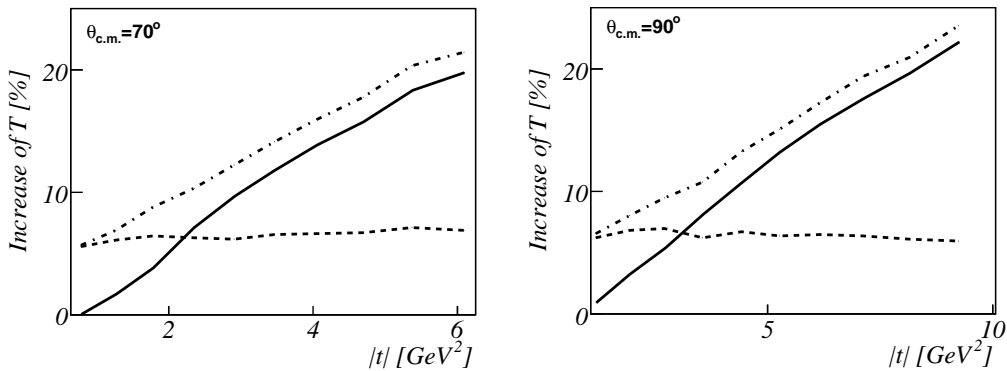}
\caption{The $\mid t \mid$-dependence of the relative increase of the
nuclear transparency due to SRC and CT effects. We consider the
$^4\text{He}(\gamma,p\pi^-)$ reaction at $\theta_{\text{c.m.}}=70^o$
(left panel) and $90^o$ (right panel) and kinematic conditions from
Table~\ref{tab:kinphoto}. The baseline result is the
RMSGA calculation.  The solid (dashed) curve includes the effect of CT
(SRC). The dot-dashed line is the combined effect of CT+SRC.}
\label{fig:photoincrease}
\end{figure}

Fig.~\ref{fig:photocompare70} also shows that the $^4\text{He}$ nucleus
is more transparent for pion emission than for proton emission.  This
can be partially attributed to the lower pion total cross sections.
As pointed out in Fig.~\ref{fig:CT} the larger formation length, and
corresponding bigger reduction of the effective cross section make
that the CT effect is larger for pions than for protons. In
Fig.~\ref{fig:photoincrease} the computed increase in the nuclear
transparency caused by CT and SRC mechanisms is shown as a function of
$|t|$.  One observes that SRC mechanisms increase the nuclear
transparency by about 5\%.  As there is no direct dependence on the
hard scale, the increase is almost independent of $|t|$.  The CT
phenomenon, on the other hand, shows a linear rise from almost 0 to
over 20\% at the largest values of $|t|$.  For $- t \le 2.5 $ ~GeV$^2$
the predicted effect of SRC is larger than the increase induced by the
CT mechanism. The SRC decrease the slope in the $ - t $-dependence of
the CT phenomenon.  Indeed, the SRC induces holes in the nuclear
density in the direct neighborhood of the interaction point (see
Fig.~\ref{fig:SRC}) where the CT effects are largest.  At high $|t|$
the short-range correlations have a modest impact on the magnitude of the CT
effects.  Our investigations show that by studying the hard scale dependence of
the transparency the CT-related mechanisms can be clearly separated from the
SRC ones.

\begin{figure}[htb]
\centering
\includegraphics[width=14cm]{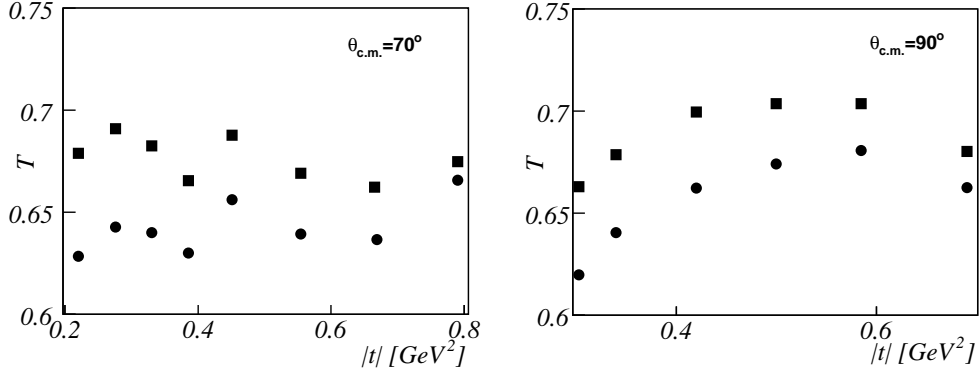}
\caption{Comparison between the RMSGA (squares) and ROMEA (circles) description
of the nucleon transparency of the $^4\text{He}(\gamma,p\pi^-)$ reaction for
kinematics at $\theta_{\mathrm{c.m.}}=70^o$ (left panel)
and $90^o$ (right panel).  Neither CT nor SRC effects were included in the
calculations.}
\label{fig:lowE}
\end{figure}

In the search of phenomena like CT in transparency studies, it is of the
utmost importance to possess robust and advanced calculations based on
concepts from traditional nuclear physics.  Thereby, one of the major
sources of uncertainty stem from the description of FSI mechanisms.  In
our eikonal model, we can either use optical potentials (ROMEA) or a
Glauber framework (RMSGA).  In some kinematic region of moderate hadron
momenta both approaches can be used \cite{Lava:2004zi}.  As they adopt very
different underlying assumptions, we consider a comparison between the
predictions of the two approaches as a profound test of the
trustworthiness of either approach.  We computed the transparency of the
$^4\text{He}(\gamma,p\pi^-)$ reaction for kinematics at
$\theta_{\mathrm{c.m.}}=70^o$ and $90^o$ with ejected proton momenta ranging
from 500 MeV/c to 1 GeV/c.  As can be appreciated from Fig. \ref{fig:lowE},
both descriptions yield a similar shape, but the RMSGA calculations are
consistently larger by about 5\%.  At higher nucleon momenta though, the
difference is at the order of a few percent.

\subsection{Pion electroproduction}
The E01-107 collaboration at Jefferson Lab has measured the nuclear 
transparency for the pion electroproduction process on $\text{H}$,
$^{12}\text{C}$, $^{27}\text{Al}$, $^{64}\text{Cu}$ and $^{197}\text{Au}$. 
Measurements
were done for the kinematics listed in Table~\ref{tab:electrokin}.  In
all the measurements the pion is detected in a relatively narrow cone
about the momentum transfer.  We have performed calculations for all
target nuclei.  The transparency is defined as
\begin{equation}\label{eq:elektrotransp}
T=\frac{\sum_{\alpha}\int d\omega Y(\omega) \int_{\Delta^3p_m}d\vec{p}_m 
 \left(\frac{d^8\sigma}{d\Omega_{e'}dE_{e'}dE_\pi d\Omega_\pi
d\Omega_N}\right)_{\text{RMSGA}}}
{\sum_{\alpha}\int d\omega Y(\omega)\int_{\Delta^3p_m}d\vec{p}_m 
 \left(\frac{d^8\sigma}{d\Omega_{e'}dE_{e'}dE_\pi d\Omega_\pi
d\Omega_N}\right)_{\text{RPWIA}}}\,.
\end{equation}
The integration over $\omega$ takes into account the spread in energy of the 
virtual photon in the experiment and weighs each point with the reconstructed 
yield $Y(\omega)$ \cite{Clasie}. The quantity $\Delta^3p_m$ specifies the phase-space of the missing
momentum and is determined by the condition $\mid p_m\mid \leq 300$
MeV/c and the experimental cuts and detector acceptances.  Accordingly, the
final neutron is extremely slow and we have assumed that it possesses a
transparency of one. A cut of 100 MeV was placed on the missing mass of the
final state. The
intranuclear attenuation effects on the ejected pion are again
computed with the RMSGA model.  We use a parametrization provided by
the E01-107 collaboration for the free electroproduction cross section in
Eq.~(\ref{electrocrosssection}) \cite{Horn,Clasie}.

\begin{table}[htpb]
    \begin{center}
        \begin{tabular}{lrrrrr}
            \hline
            \hline
	    $Q^2$ & $E_e$ & $\theta_e$ & $E_{e'}$ & $p_\pi$ & $\theta_\pi$\\
	    \hline
	    1.10 & 4021 & $27.76^o$ & 1190 & 2793 & $10.58^o$ \\
	    2.15 & 5012 & $28.85^o$ & 1730 & 3187 & $13.44^o$ \\
	    3.00 & 5012 & $37.77^o$ & 1430 & 3418 & $12.74^o$ \\
	    3.91 & 5767 & $40.38^o$ & 1423 & 4077 & $11.53^o$ \\
	    4.69 & 5767 & $52.67^o$ & 1034 & 4412 & $9.09^o$ \\
	    \hline
            \hline
        \end{tabular}
    \end{center}
\caption{Central values of $Q^2$ ($\text{GeV}^2$), incoming electron
energy $E_e $(MeV), electron scattering angle $\theta _e$ (degrees), scattered
electron energy $E_{e'}$ (MeV), ejected pion momentum  $p_\pi$ (MeV)
and ejected pion angle (degrees) for the kinematics of the Jefferson Laboratory
experiment
E01-107.  Angles are measured relative to the incoming electron beam.}
\label{tab:electrokin}
\end{table}

\begin{figure}[htb]
\centering
\includegraphics[width=14cm]{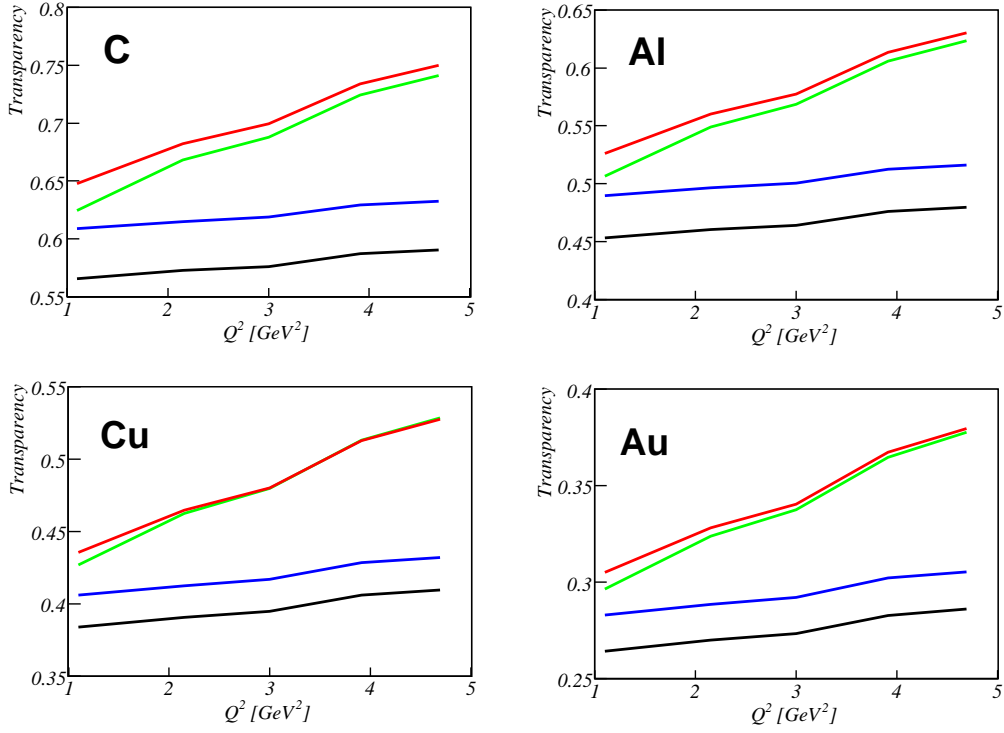}
\caption{(Color online) The $Q^2$-dependence of the nuclear
  transparency for the $A(e,e' \pi^+)$ process in $^{12}\text{C}$,
  $^{27}\text{Al}$, $^{63}\text{Cu}$ and $^{197}\text{Au}$. The black
  and green curves are  RMSGA and RMSGA+CT calculations respectively. The blue and 
  red line are RMSGA+SRC and RMSGA+SRC+CT results.}
\label{fig:electrotransp}
\end{figure}

\begin{figure}[htb]
\centering
\includegraphics[width=7cm]{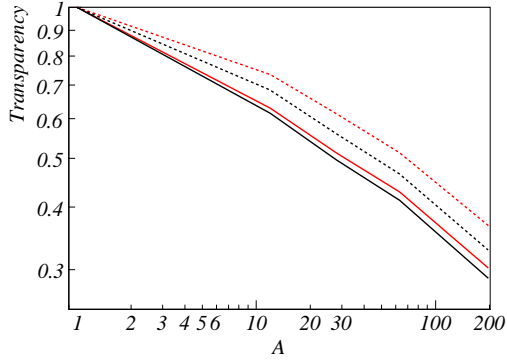}
\caption{(Color online) A-dependence of the transparency for the $A(e,e'\pi^+) $ process at
$Q^2=1.1\, \text{GeV}^2$ (black) and $Q^2=4.69\, \text{GeV}^2$ (red).
  The solid curves denote RMSGA+SRC results.  The dashed lines are
  RMSGA+CT+SRC calculations.}
\label{fig:electroAdep}
\end{figure}

Fig.~\ref{fig:electrotransp} presents the results from our
transparency calculations for the electroproduction reaction.  The
RMSGA calculations show a modest increase over the $Q^2$ range. This
behavior finds a simple explanation in the $p_{\pi}$ dependence of the
$\sigma _{\pi^{+} p}^{\text{tot}}$ of Fig.~\ref{fig:pioncross}. The
results contained in Fig.~\ref{fig:electrotransp} cover a range in
pion momenta given by $2.8 \le p _{\pi} \le 4.4$~GeV. In this range,
$\sigma _{\pi^{+} p}^{\text{tot}}$ displays a soft decrease, which
reflects itself in a soft increase of the nuclear transparency.  The RMSGA+CT
transparencies are again about $5\%$ larger than the RMSGA ones.  The
RMSGA+CT shows a strong $Q^2$ dependence with CT-related enhancements
up to $20\%$ at the highest energies.  The evolution of the
A-dependence of the transparency is shown in Fig.
\ref{fig:electroAdep}.  One observes that the addition of CT to the
calculation adds more curvature and that this increases with higher
$Q^2$. Finally, we compare our model calculations with the results
from the semi-classical model of Ref. \cite{Larson:2006ge}.  The transparency
is plotted as function of $\vec{k}=\vec{p}_\pi-\vec{q}$.  As in the
photoproduction calculations \cite{Cosyn:2006vm}, our results again
turn out to be higher by a few percent.

\begin{figure}[htb]
\centering
\includegraphics[width=7cm]{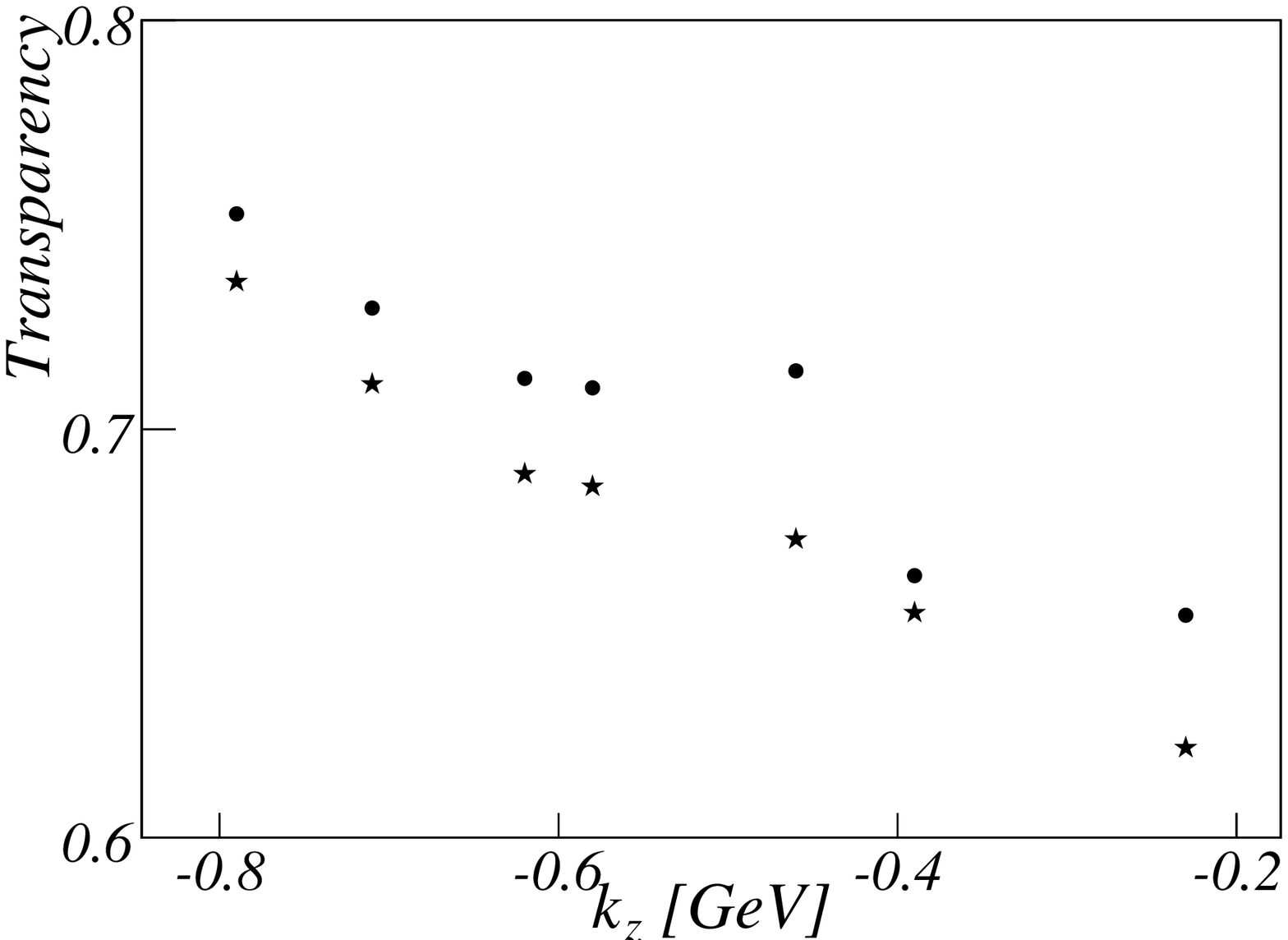}
\caption{Nuclear transparency results for $^{12}\text{C} (e,e' \pi^+)$
 versus the $z$ component of $\vec{k}=\vec{p}_\pi-\vec{q}$ for kinematics
corresponding to data points of
 the JLab experiment of Ref. \cite{Clasie:2007gq}.  The circles are RMSGA+CT predictions,
 whereas the stars are from the semi-classical calculations of
 Ref.~\cite{Larson:2006ge}.}
\label{fig:Millercompare}
\end{figure}

\section{Conclusion} \label{sec:conclusion}
We have outlined a relativistic framework to compute nuclear
transparencies in exclusive $A(\gamma,N \pi)$ and $A(e,e' N \pi)$
reactions.  For the bound states, the model uses relativistic
mean-field wave functions. At sufficiently high nucleon and pion
energies, the intranuclear attenuation on the ejected particles can be
computed with a relativistic version of the Glauber model. At lower
ejectile energies, the framework offers the flexibility to use optical
potentials. For nucleon momenta where both approaches can be applied, the
Glauber and optical-potential based calculations predict nucleon
transparencies in $^4\text{He}$ which follow similar trends.  The
differences in the magnitude of the transparency is smaller than 5\% and
shrinks with nucleon momentum.  Our RMSGA predictions for the pion
transparencies are in reasonable agreement with the semi-classical results of
Larson, Miller and Strikman.  Both models predict similar trends, with the
RMSGA predictions being systematically $\approx 5\%$ higher.
This provides support that the baseline nuclear-physics transparencies 
can be computed in a rather model-independent fashion.  Extension of our relativistic and 
quantum mechanical photoproduction calculations up to
energies accessible in the JLab 12 GeV upgrade show an increase of the
transparency up to 20\% at the highest energies due to color transparency.
Transparencies are also enhanced through the inclusion of
SRC effects in the calculations.  This yields an increase of about 5\%,
independent of the hard scale.  Accordingly the SRC and CT mechanisms can be
clearly separated.

\end{document}